\documentclass[twocolumn]{aa}

\usepackage[T1]{fontenc}
\usepackage{lmodern}
\usepackage[nolist]{acronym}
\usepackage{graphicx}
\usepackage{txfonts}
\usepackage{hyperref}
\usepackage{booktabs}
\usepackage{xcolor}
\usepackage[utf8]{inputenc}
\usepackage{lscape}
\usepackage{placeins}
\usepackage{xurl}
\usepackage{natbib}
\usepackage{enumitem}
\usepackage{placeins}

\bibpunct{(}{)}{;}{a}{}{,}

\graphicspath{{fig/}}

\newcommand{\XMM}{{\sl XMM-Newton}}
\newcommand{\xmmiiathena}{{\sl XMM2ATHENA}}
\newcommand{\pn}{{pn}}
\newcommand{\mos}{{MOS}}
\newcommand{\mosi}{{MOS1}}
\newcommand{\mosii}{{MOS2}}
\newcommand{\zpltb}  {{\em zpltb}}
\newcommand{\apec}   {{\em APEC}}
\newcommand{\bbpl}   {{\em bbpl}}
\newcommand{\bremss} {{\em bremss}}

\newcommand{\mylowt}         {{\sl lowT}}
\newcommand{\myhight}        {{\sl highT}}
\newcommand{\myabsagn}       {{\sl absAGN}}
\newcommand{\myunabsagn}     {{\sl unabsAGN}}
\newcommand{\myundetagn}     {{\sl undetAGN}}
\newcommand{\mylognhxxiii}   {{\sl logNH23}}
\newcommand{\myabsunabsundet}{{\sl logNH90\%}}

\newcommand{\mylogfx}{\log{(f_{\rm X} / {\rm erg\,cm^{-2}\,s^{-1}})}}
\newcommand{\mylognh}{\log{(N_{\rm H} / {\rm cm^{-2}})}}
\newcommand{\myloglx}{\log{(L_{\rm X} / {\rm erg\,s^{-1}})}}

\newcommand{\dFourOne}{180\,118}
\newcommand{\dFourTwo}{30\,326}

\newcommand{\dFourToFit}{51\,166}
\newcommand{\dFourToFitAGN}{35\,538}
\newcommand{\dFourToFitStar}{14\,308}
\newcommand{\dFourToFitXRB}{1\,091}
\newcommand{\dFourToFitCV}{229}

\newcommand{\dFourDetThereZero}{24\,402}
\newcommand{\dFourDetThereOne}{5\,579}
\newcommand{\dFourDetThereTwo}{21\,185}

\begin{document}

\title{%
   Harnessing the \XMM{} data: X-ray spectral modelling of 4XMM-DR11
   detections and 4XMM-DR11s sources%
}

\titlerunning{X-ray spectral modelling of 4XMM-DR11}

\authorrunning{Viitanen et al.}

\author{%
  A.~Viitanen\inst{1,2,3}
  \and G.~Mountrichas\inst{3}
  \and H.~Stiele\inst{4,3}
  \and F.~J.~Carrera\inst{3}
  \and A.~Ruiz\inst{5}
  \and J.~Ballet\inst{6}
  \and A.~Akylas\inst{5}
  \and A.~Corral\inst{3}
  \and M.~Freyberg\inst{7}
  \and A.~Georgakakis\inst{5}
  \and I.~Georgantopoulos\inst{5}
  \and S.~Mateos\inst{3}
  \and C.~Motch\inst{8}
  \and A.~Nebot\inst{8}
  \and H.~Tranin\inst{9,10,11}
  \and N.~A.~Webb\inst{12}
}

\institute{%
  INAF–Osservatorio Astronomico di Roma, via Frascati 33, 00040 Monteporzio Catone, Italy
  \and Department of Physics, University of Helsinki, PO Box 64, FI-00014 Helsinki, Finland
  \and Instituto de F\'{\i}sica de Cantabria (CSIC-Universidad de Cantabria), Avenida de los Castros, 39005 Santander, Spain
  \and Juelich Supercomputing Centre, Forschungszentrum Juelich GmbH, 52425 Juelich, Germany
  \and Université Paris Saclay and Université Paris Cité, CEA, CNRS, AIM, F-91191 Gif-sur-Yvette, France
  \and Institute for Astronomy Astrophysics Space Applications and Remote Sensing (IAASARS), National Observatory of Athens, Ioannou Metaxa \& Vasileos Pavlou, Penteli 15236, Greece
  \and Max-Planck-Institut für extraterrestrische Physik, Giessenbachstraße 1, Garching 85748, Germany
  \and Université de Strasbourg, CNRS, Observatoire astronomique de Strasbourg, UMR 7550, Strasbourg 67000, France
  \and Institut de Ci\`{e}ncies del Cosmos (ICCUB), Universitat de Barcelona (UB), c. Martí i Franqu\`{e}s, 1, 08028 Barcelona, Spain
  \and Departament de F\`{i}sica Qu\`{a}ntica i Astrofísica (FQA), Universitat de Barcelona (UB), c. Mart\`{i} i Franqu\`{e}s, 1, 08028 Barcelona, Spain
  \and Institut d’Estudis Espacials de Catalunya (IEEC), c/ Esteve Terradas,1, Edifici RDIT, Despatx 212, Campus del Baix Llobregat UPC - Parc Mediterrani de la Tecnologia, 08860 Castelldefels, Spain
  \and IRAP, Université de Toulouse, CNRS, UPS, CNES, 9 Avenue du Colonel Roche, BP 44346, 31028 Toulouse Cedex 4, France
}

\date{Received April 23, 2025; accepted September 24, 2025}

\abstract{%
  The \XMM{} X-ray observatory has played a prominent role in astrophysics,
  conducting precise and thorough observations of the X-ray sky for the past
  two decades. The most recent iteration of the \XMM{} catalogue, 4XMM, and
  one of its latest data releases, DR11, mark significant improvements over
  previous \XMM{} catalogues, serving as a cornerstone for comprehending the
  diverse inhabitants of the X-ray sky. In this investigation, we employ
  X-ray detections and spectra extracted from the 4XMM-DR11 catalogue,
  subjecting them to fitting procedures using simple models. Our study
  operates within the framework of the \xmmiiathena{} project,
  which focuses on developing state-of-the-art methods that exploit
  existing \XMM{} data. In
  this study, we introduce and publicly release four catalogues containing
  measurements derived from X-ray spectral modelling of sources. The first
  catalogue encompasses outcomes obtained by fitting an absorbed power law
  model to all the extracted spectra for individual detections within the
  4XMM-DR11 dataset. The second catalogue presents results obtained by
  fitting both an absorbed power law and an absorbed blackbody model to all
  unique physical sources listed in the 4XMM-DR11s catalogue, which documents
  source detection results from overlapping \XMM{} observations. For the
  third catalogue we use the five band count rates derived from the pipe line
  detection of X-ray sources to mimic low resolution spectra to get a rough
  estimate of the spectral shape (absorbed power-law) of all 4XMM-DR11
  detections. In the fourth catalogue, we conducted spectral analyses for the
  subset of identified sources with extracted spectra, employing various
  models based on their classification into categories such as \acp{AGN},
  stars, \aclp{XRB}, and \aclp{CV}. Finally, the scientific potential of these
  catalogues is highlighted by discussing the capabilities of optical and
  mid-infrared colours for selecting absorbed \acp{AGN}\@.%
}


\maketitle

\acresetall

\section{Introduction}

The study of X-rays from celestial sources opens a gateway to a realm of
high-energy astrophysical phenomena, enabling us to delve into the nature of
some of the Universe's most extreme and time-variable objects. The \XMM{}
X-ray observatory \citep{Jansen2001} is a cornerstone mission of the European
Space Agency's (ESA) Cosmic Vision programme. With its large field of view
(about 30~arcmin diameter), a PSF with a full width at half maximum (FWHM) of
$\sim6\arcsec$ and a half-energy width (HEW) of $\sim15\arcsec$, along with a
large collecting area of $4\,500\,{cm}^2$ at $1\,{\rm keV}$ (the largest of all
current missions), it is an ideal tool for performing surveys and spectral
analysis to investigate the physical properties of cosmic X-ray sources. The
\XMM{} catalogues can be considered the European counterpart to the \ac{CSC},
whose latest release is version 2.1
\citep{Evans2024}\footnote{\url{https://cxc.cfa.harvard.edu/csc/}}. The
\ac{CSC} provides high-angular-resolution source data, including spectral
properties\footnote{\url{https://cxc.cfa.harvard.edu/csc/columns/spectral_properties.html}},
and complements \XMM{} in terms of angular resolution and survey depth.

To harness \XMM{}'s data, the \XMM{} Survey Science Centre
\citep[XMM-SSC;][]{Watson2001}, a collaboration of ten European institutes in
conjunction with the \XMM{} Science Operations Centre (SOC), has developed the
\acused{SAS} \acl{SAS} (\acs{SAS}; \citealp{Gabriel2004}) software suite. The
\ac{SAS} enables the reduction and analysis of \XMM{} data, supported by a
dedicated pipeline for standardised processing of the science data, ultimately
leading to the creation of catalogues containing information on X-ray and
optical/UV sources \citep{Page2012,Traulsen2020,Webb2020}. Catalogues serve as
indispensable resources for a wide range of scientific inquiries, providing
homogeneous datasets for classes of objects and unveiling previously unknown
sources.

The X-ray detection catalogues created from the observation data of the three
camera systems (one \pn{} and two \mos{}) of the European Photon Imaging Camera
\citep[EPIC;][]{Turner2001} have been identified as 1XMM, 2XMM, and 3XMM, each
representing a successive iteration marked by data releases referred to as `DR'
in conjunction with the catalogue number. The latest version of the XMM
catalogue, 4XMM \citep{Webb2020}, and its latest DR (DR11 at the time of
starting this work) incorporates many improvements with respect to previous
\XMM{} catalogues and serves as a cornerstone for understanding the X-ray sky's
diverse inhabitants. The 4XMM-DR11 catalogue represents the culmination of over
two decades of meticulous X-ray observations by the \XMM{} satellite.

In this study, we utilise the X-ray detections and spectra extracted from the
4XMM-DR11 catalogue, and the unique X-ray sources from the 4XMM-DR11s
catalogue, and subject them to automated fitting procedures employing both
simple and physically motivated models. Our investigation is carried out within
the framework of the \xmmiiathena{} \citep{Webb2023a} project, which is
dedicated to developing novel methodologies for harnessing the existing \XMM{}
data and preparing for its seamless integration with forthcoming
\textsl{NewAthena} observations. In particular, we focus on automated X-ray
spectral fitting pipelines and catalog-level modelling approaches that enable
population-wide analysis across millions of detections. This endeavor
encompasses the incorporation of multi-wavelength and multi-messenger data,
providing a comprehensive approach to unravelling the intricate cosmic phenomena
captured by these advanced observatories. Specifically, we aim to reveal the
X-ray spectral properties of different classes of X-ray sources, such as
\acp{AGN}, \acp{XRB}, \acp{CV}, and stars.

Direct antecedents for this work are XMMFITCAT \cite{Corral2015} and
XMMFITCAT-Z \cite{Ruiz2021} (R21), which provided spectral fits for $>114\,000$
detections from 3XMM-DR4 and $22\,677$ identified sources from 3XMM-DR6 (mostly
\acp{AGN}), respectively. With respect to them, we are using later versions of
the \XMM{} catalogues with more detections and sources, and hence also more
extracted spectra, and updated identifications and photometric redshifts. On
the other hand, we are using a reduced set of spectral models, and fits in a
single spectral band. More detailed discussions of the differences will be
included in Sects.~\ref{sec:C1_analysis} for XMMFITCAT
and~\ref{sec:C4_analysis} for XMMFITCAT-Z\@.

The paper is organised as follows: Sect.~\ref{sec:data} offers a concise
overview of the 4XMM-DR11 catalogues and outlines the sources encompassed in
the four catalogues we generated. Sect.~\ref{sec:models} provides insights into
the spectral models employed to fit X-ray spectra in each of the four
catalogues, along with details on source classification and photometric
redshift calculation. In Sect.~\ref{sec:results}, we present measurements of
the primary properties of the sources and conduct a comparative analysis across
the four catalogues. Sect.~\ref{sec:science} showcases a scientific application
utilizing one of the generated catalogues. Finally, Sect.~\ref{sec:conclusions}
summarises the key findings of this study.

\section{Description of the parent and generated catalogues}
\label{sec:data}

In this section, we provide a brief description of the parent catalogues we
used in our analysis. We also describe in detail the four catalogues we
compiled.

\subsection{The parent catalogues}
\label{sec:data_parent}

The X-ray sources utilised in our analysis were extracted from the 4XMM-DR11
catalogue \citep{Webb2020}, which is based on 12\,210 \XMM{} EPIC observations
and contains 895\,415 detections surpassing a statistical detection likelihood
threshold of six. These correspond to 602\,543 unique sources, approximately
19\% of which have more than one detection. The catalogue includes both
point-like and extended sources, with extent parameters considered reliable up
to a maximum extent of 80\arcsec. The median fluxes in the total (0.2–12 keV),
soft (0.5–2 keV), and hard (2–12 keV) bands are $\sim 2.3 \times 10^{-14}$,
$5.2 \times 10^{-15}$, and $1.2 \times 10^{-14}$ erg cm$^{-2}$ s$^{-1}$,
respectively, with 16th and 84th percentile ranges of [${\sim}5 \times 10^{-15},
{\sim}8 \times 10^{-14}$], [${\sim}1.4 \times 10^{-15}, {\sim}2.1 \times
10^{-14}$], and [${\sim}3.2 \times 10^{-15}, {\sim}4.1 \times 10^{-14}$].
Spectra are extracted for detections with more than 100 EPIC net counts in the
$0.2-12\,{\rm keV}$ band (see \citealt{Traulsen2020,Webb2020} for details on
extraction, background modelling, and the treatment of extended emission).

For the detections with more than $100$ $0.2-12\,{\rm keV}$ net EPIC counts one
spectrum per available camera (\pn{}, \mosi{}, and \mosii{}) were extracted
from the source region (including source and background counts) and from a
nearby region devoid of sources \cite[including only background counts:
see][for details]{Webb2020}. We refer to the first as the source spectra and to
the second as the background spectra.

An additional independent catalogue, termed 4XMM-DR11s, has been concurrently
compiled by the \XMM{} SSC, with the `s' denoting `stacked'. This catalogue
provides a record of source detection from 8\,274 overlapping \XMM{}
observations. The 4XMM-DR11s contains $1\,488$ stacks. To achieve simultaneous
source detection on these overlapping observations, individual events were
adjusted in position based on the outcomes of the preceding {\tt catcorr}
positional correction applied to the entire image during the processing of
4XMM-DR11. This adjustment resulted in a noticeable enhancement in the
positional accuracy when conducting stacked source detection. All sources
identified through stacked source detection are documented in 4XMM-DR11s,
including those originating from image areas where only a single observation
contributes. It is worth noting that there may be disparities between the same
sources listed in 4XMM-DR11 and DR11s, as their input event lists are treated
differently: in DR11s, stacked source detection is performed after correcting
and combining multiple overlapping observations, which can lead to improved
source positions, refined source parameters (e.g., extent, flux), and in some
cases, detection of fainter sources not visible in individual observations
\citep{Traulsen2020}. The stacked catalogue includes $358\,809$ sources, of
which $275\,440$ have several contributing observations.

In the context of this paper we would like to emphasise the differences between
detections, sources and stacked sources: the same physical source can give rise
to several detections in 4XMM-DR11, one for each time \XMM{} pointed in its
direction. Each of these detections is represented by a unique detection
identifier {\tt DETID}. Within 4XMM-DR11 the unique physical sources have been
determined, assigning to each one of them a unique source identifier {\tt
SRCID}, so several detections can have the same source identifier. On the other
hand, each entry in the stacked catalogue 4XMM-DR11s corresponds to a unique
physical source, with their unique identifier also named {\tt SRCID}. Their
correspondence with the 4XMM-DR11 sources and detections (when found) is
included in the stacked catalogue in additional rows, containing their
detections and source identifications as {\tt DETID\_4XMMDR11} and {\tt
SRCID\_4XMMDR11}, respectively.

\subsection{Description of the compiled catalogues}
\label{sec:catalogues_analysis}

We generated four catalogues using the 4XMM-DR11/DR11s parent catalogues. Below
is an overview of these datasets.

\subsubsection{Modelling the extracted spectra of 4XMM-DR11}
\label{sec:C1_analysis}

In the first catalogue (catalogue C1 hereafter), we present the results from
fitting an absorbed power law model (detailed in the next section) to all
extracted spectra in the 4XMM-DR11. We furnished the parameter values that
yield the best fit as well as their associated confidence intervals. To
expedite the execution of spectral fits, we merged all source and background
spectra from the same detection and camera within the same observation using
the \ac{SAS} task \textit{epicspeccombine}. This approach ensures that, for each
detection, we ended up with a maximum of one \pn{} spectrum and one \mos{}
spectrum.

For the spectral fitting and modelling procedures, we employed the analysis
software Sherpa 4.9.1 \citep{Freeman2001} and the \ac{BXA} tool
\citep{Buchner2014}. The \ac{BXA} tool facilitates the connection between XSPEC
\citep{Arnaud1996} and the nested sampling package \textit{UltraNest}
\citep{Buchner2021}. We assigned uninformative priors to each parameter within
the model and explored the entire parameter space using equal-weighted sampling
points, conducted via the \textit{MLFriends} algorithm \citep{Buchner2019},
which is integrated within \textit{UltraNest}.

We perform spectral fitting using the Cash statistic \citep{Cash1979}, which is
well suited for Poisson-distributed data, especially in the low-count regime.
The fitting procedure is as follows: First, we merged the spectra of all
exposures for the same EPIC camera type (\pn{} and \mos{}) using the \ac{SAS} task
\textit{epicspeccombine}, resulting in a maximum of one \pn{} and one \mos{}
spectrum for each detection. Second, the background spectra for each camera
(\pn{} and \mos{}) are grouped to ensure a minimum of one count per bin. These
background spectra are then fit using an empirical model tailored to each
camera (see Sect.~\ref{sec:back_models}). Fits with probabilities $p < 0.01$
(see Sect.~\ref{sec:back_models}) are rejected and not used for further
analysis. These $p$-values (\texttt{pval\_bg\_pn}, \texttt{pval\_bg\_mos}) are
reported in our catalogues, and the \texttt{det\_use} flag depends on their
outcome.

Then, we bin the source$+$background spectra similarly ($\geq 1$ count per
bin), preserving the Poisson nature of the data. Finally, we fit the combined
spectra for both cameras with a source+background model, where the background
model parameters (except the normalisation) are fixed to the best-fit values
obtained in the first step. This ensures consistency and prevents overfitting.
For background spectra, the typical number of bins per camera ranges from 20 to
60, depending on the exposure and source brightness, with more than 90\% of
cases having at least 10 bins. Background spectra with zero counts are excluded
from the analysis.

Although joint fitting of the source and background spectra with all components
free is often preferred for propagating uncertainties, we opted to model the
background separately and fix its shape during the source$+$background fit.
This decision was motivated by the complexity of the empirical \XMM{}
background, which includes numerous components with many free parameters, and
by the need to ensure robust convergence in a fully automated pipeline. By
fitting the background first, we allow better control over the model components
and avoid degeneracies with the source model \citep[see e.g.][]{Buchner2014}.

Out of the $895\,415$ detections listed in the 4XMM-DR11 catalogue, $319\,565$
of them, originating from a total of $11\,907$ observations, contain
significant count numbers that qualify them for automated spectral extraction
within the processing pipeline \citep{Webb2023a}. For $390$ detections ($\sim
0.1\%$) the automated definition of a background extraction region of at least
one camera failed and the resulting background spectrum has no counts. If we
also demand that a detection has more than zero net counts in each contributing
camera (\pn{} and/or \mos{})\footnote{No background subtraction is done at the
spectral fitting stage, this filtering is done only for quality purposes} a
further $4\,435$ detections ($\sim 1.5\%$) are excluded. This results in
$314\,352$ detections that constitute what we call the \textit{Good sample}. Out
of these detections, $245\,484$ ($\sim 80\%$) gave an acceptable fit for the
background model (i.e.\ $\chi^2$ $p$-value > 0.01, see
Sect.~\ref{sec:back_models}) and $232\,816$ ($73.8\%$) of them also gave an
acceptable fit for the source model. These sources comprise, what we call the
\textit{Good fit} sample (see Table~\ref{tab:table_samples}). Among these,
$100\,237$ detections (making up 43.1\%) are present in both cameras,
$135\,342$ detections (constituting 58.2\%) exclusively stem from the \pn{}
camera, and $73\,986$ detections (representing 31.8\%) solely arise from the
\mos{} camera.

\citet{Corral2015} in XMMFITCAT provided fit results for $>114\,000$
detections, corresponding to $\sim 78\,000$ unique sources, using three bands
(soft: $0.5-2\,{\rm keV}$, hard: $2-10\,{\rm keV}$, full: $0.5-10\,{\rm keV}$)
and six spectral models (three simple and three more complex ones, the latter
only applied to sources with more than 500 counts). They used the default Xspec
algorithm (Levenberg-Marquardt) to find the best-fit values for each model
parameter, with some optimisations included in their scripts to compensate for
its tendency to find local rather than global minima. We provide fit results
for $319\,565$ detections, corresponding to $213\,154$ unique sources, using
\ac{BXA}, which makes a thorough search of the parameter space using \textit{UltraNest},
making it much better suited to find global minima. While they fitted the
background-subtracted spectrum using the Cash statistic, we fitted first the
background file using \ac{BXA} and an empirical model, and then we fitted the
source+background spectrum with the background model parameter values fixed to
the best-fit background-only values, apart from the normalisation. A further
refinement is that our \ac{GoF} calculation accounts for the fact that the data
and the simulations used to estimate the \ac{GoF} are not independent, which is
not taken into account using {\tt goodness} within Xspec, as they did. On the
other hand, the sheer number of fits constrained us to use a single band
($0.2-12\,{\rm keV}$) and a single model (an absorbed power-law, also included
in XMMFITCAT).

\begin{table}
  \centering
  \caption{%
    Number of detections (C1, C3) or sources (C2, C4) included in each one of
    the four compiled catalogues.%
  }
  \begin{tabular}{lrrr}
    \toprule
    Catalogue & Total      & \textit{Good sample} & \textit{Good fit} sample  \\
    \midrule
    C1        & $319\,565$ & $314\,352$         & $232\,816$              \\
    C2        &  $32\,622$ &  $30\,325$         &  $23\,426$ ($15\,352$)  \\
    C3        & $895\,415$ & $419\,118^*$        & $400\,390$              \\
    C4        &  $51\,166$ &  $50\,956$         &  $41\,181$              \\
    \bottomrule
  \end{tabular}
  \tablefoot{%
    For the definitions of \textit{Good} and \textit{Good fit} samples see the
    text for each catalogue. For the \textit{Good fit} sample of C2, the values
    correspond to the two separate models: the power-law and the blackbody (in
    parenthesis), respectively. $^*$ This corresponds to the \textit{Clean
    sample}, see text.
    }%
  \label{tab:table_samples}
\end{table}

\subsubsection{Modelling the stacked spectra of 4XMM-DR11s}
\label{sec:C2_analysis}

In the second catalogue that we release (catalogue C2 hereafter), we fitted
both an absorbed power law model and an absorbed blackbody model to sources
from the 4XMM–DR11s catalogue. Using two models for the full set of 4XMM-DR11
detections (see C1 above) was not feasible due to the significantly larger
number of sources involved and the associated computational cost.

The 4XMM-DR11s catalogue contains 60\,720 unique sources associated with
135\,612 detections with extracted spectra. Among these, 27\,640 sources are
associated with only a single detection. For such sources, an absorbed
power-law model has already been applied in C1, where individual detections
were modeled. Since the aim of C2 is to exploit the additional information from
multiple detections by stacking them, we do not re-fit sources with only one
detection in this catalogue. These sources therefore remain part of C1 only,
and are not included in C2. This ensures that the added complexity of C2,
including model comparison and stacked spectra, is applied only to cases where
multi-epoch data provides additional value.

Following a methodology similar to that used in the case of C1, we excluded
detections where at least one camera's background spectrum contained no counts
or where the net counts in at least one camera were less than zero. For the
remaining sources with just one contributing detection after this step, we
already possess spectral information from C1. Consequently, a total of 458
sources were omitted from C2, leaving us with $32\,622$ remaining sources. For
these, we computed the \ac{S/N} for each individual detection (by summing the
counts from \pn{} and \mos{} in the full band). We then sorted the detections
in descending order of \ac{S/N} and calculated a cumulative S/N (cS/N) for each
detection, incorporating all detections with an equal or higher \ac{S/N}\@.
Combining spectra from multiple observations results in an average spectrum
that represents the time-averaged source properties. This approach is
appropriate for most sources, especially given that our spectral models are
relatively simple and are not designed to capture detailed spectral evolution.
While strong variability could introduce complexities in interpreting the
averaged parameters, the stacked spectrum remains a valid representation of the
source’s mean behavior over the combined epochs.

Out of the $19\,081$ sources with only two contributing detections, we used
both detections if the cS/N increased when including the second,
lower \ac{S/N} spectrum (this was the case for $16\,959$ sources). In contrast, for
the remaining $2\,122$ sources, only the first detection was considered, as
their spectral properties are already covered by analysis followed for the
first catalogue (i.e., C1) and, therefore, not included in this part. We also
excluded $173$ sources from C2 for which the detection with the highest
individual \ac{S/N} coincided with the highest cS/N, as only one
spectrum would contribute.

For the sources with more than two contributing detections ($13\,541-173 =
13\,368$ sources), we introduced a selection criterion based on the relative
range (rr) of the cS/N to optimise the number of observations that
are included in the spectral fitting\@. This relative range is calculated as
the difference between the cS/N and the maximum individual value, divided by
the average of the individual values:
\begin{equation}
    {\rm rr}_n =
    \frac{{\left( \rm cS/N \right)}_n-\mathrm{max}({\rm S/N})}
         {1/n\sum_{i=1}^n {\left( \rm S/N \right)}_i}
\end{equation}

Our aim was to identify the detection where the cS/N reached its
maximum or where adding further detections did not improve it significantly
anymore. Especially in the latter case, it is better to use the relative rather
than the absolute range, as we can then define a negligible increase by a fixed
value that is applicable to all sources (see below). This last detection, along
with all detections possessing a higher individual \ac{S/N} than this last one,
would be included in our spectral analysis of the source.

To address situations where there might be some `flickering' in the cS/N, we
imposed a limit of $>0.001$ for the increase in the relative cS/N\@. In other
words, if the change in relative cS/N between the new detection and the
previous one (with a higher individual \ac{S/N}) was $\leq 0.001$, the new
detection would not be included in the stacked spectrum. In two cases, only one
detection remained after this procedure, and consequently, these two sources
were also excluded from our study.

After implementing the aforementioned process, we were left with $30\,325$
sources with at least two contributing detections, which we refer to as the
`Good sample' (the average number of spectra used for the final stacked
spectrum per source is 3, with contributions from between 2 and 44 spectra). We
merged the spectra of all contributing detections for the same camera using the
\ac{SAS} task \textit{epicspeccombine}, resulting in a maximum of one \pn{} and one
\mos{} spectrum for each source. Specifically, $19\,973$ sources had spectra in
both cameras, $7\,311$ sources were solely observed with the \pn{} camera, and
$3\,041$ sources were exclusively obtained from the \mos{} cameras. Out of the
$30\,325$ sources, the number that meet the requirements described in the
previous section and are included in the \textit{Good fit} samples are, $23\,426$
that were fitted with a power-law model and $15\,352$ with the blackbody model.

\subsubsection{Modelling the count rates of 4XMM-DR11}
\label{sec:C3_analysis}

As part of the detection process for each observation, count rates are
determined for each \XMM{} camera in standard bands $1-5$. The five standard
bands $1-5$ correspond to energies $0.2-0.5\,{\rm keV}$, $0.5-1\,{\rm keV}$,
$1-2\,{\rm keV}$, $2-4.5\,{\rm keV}$, $4.5-12\,{\rm keV}$, respectively.

We used these detection count rates from the 4XMM-DR11 catalogue to build X-ray
spectra for the $895\,415$ detections included in the catalogue. Using the
count rates in the five defined energy bands for each EPIC camera, along with
proper response matrices (RMF and ARF, see below), we obtained a set of data
equivalent to very low resolution X-ray spectra. In other words, this technique
is roughly equivalent to extracting an X-ray spectrum and grouping it into five
bins corresponding to the 4XMM energy bands. These spectra can be fitted in the
same way as the spectra in the previous sections. The results of this analysis
are given in our third catalogue (C3 from now on). Using this method we are
able to give at least a crude estimate for the spectral parameters of all
sources in the 4XMM-DR11, even in those cases where, given the low number of
counts, a proper X-ray spectrum was not extracted.

For each of these low resolution spectra, we used as RMF the canned matrix
calculated by the \XMM{} calibration team for the corresponding camera,
epoch and mode of the
observation\footnote{\url{https://www.cosmos.esa.int/web/xmm-newton/epic-response-files}}.
As ARF matrices we calculated a set of matrices using the \textit{arfgen}
\ac{SAS}
tool, one per camera and filter (Thin, Medium and Thick). We took into account
that the count rates are already corrected for vignetting, camera efficiency,
PSF losses, bad pixels and CCD gaps, so none of these effects are included in
the ARF generation. The count rates are background-subtracted, so no background
spectra are needed. In this case, the \mos{} spectra were not merged.

In our case the likelihood probability is estimated through the $\chi^2$ value
for a set of model parameters ($\log L = -\chi^2/2$). By construction our count
rate spectra are binned, background subtracted X-ray spectra, so no other
statistic, more suited for Poisson-distributed data \citep[e.g.,][]{Cash1979},
can be employed. Note also that for sources with very low count rates some of
the bins in our spectra have less than 20 counts, and therefore a $\chi^2$
statistic is not correct from a statistical point of view \citep{Cash1979}.
Hence, be aware that our procedure only gives a quick, rough estimate of the
posterior distribution of the spectral parameters. For more rigorous results, a
proper X-ray spectral modelling should be done. In those cases where the source
is included in the C1 or C2 catalogues, we strongly recommend using those
results.

We fitted the $895\,415$ detections included in the 4XMM-DR11 catalogue,
obtaining an acceptable fit for 89.7\% of them. Since we did not include any
filtering in our selection, the catalogue can include a non negligible number
of spurious sources, detections in problematic fields or with other
observational issues. We defined a `clean' sample by selecting detections with
${\tt SUM\_FLAG} \leq 1$, ${\tt OBS\_CLASS} \leq 3$ and ${\tt EP\_8\_DET\_ML}
\geq 10$. Moreover, the spectral model we selected is reasonably flexible for
\ac{AGN} sources, but not so well suited for other X-ray populations, like clusters,
hot stars, neutron stars, \acp{XRB}, etc. In order to minimise the non-\ac{AGN}
contamination in the clean sample, we also included only sources with ${\tt
EP\_EXTENT\_ML} \leq 1$ and above the Galactic plane ($|b|>20^\circ$), where
the bulk of the stellar population is concentrated. Thus $419\,118$ detections
remain within this clean sample, with 95.5\% of them having an acceptable fit
(\textit{Good fit} sample).

\subsubsection{Modelling the classified sources of 4XMM-DR11/DR11s}
\label{sec:C4_analysis}

In the fourth catalogue (C4) we performed spectral fitting for the spectra of
sources with available classification from the \cite{Tranin2022} sample, as
described in Sect.~\ref{sec:data_class}. To generate this catalogue, we needed
to merge the C1 and C2 catalogues, avoiding multiple appearances of the same
physical source. We started by excluding from C1 all the {\tt DETID} associated
with the {\tt SRCID\_4XMMDR11} included in C2. The remaining detections from C1
were appended to the stacked sources in C2 to generate a merged catalogue with
the desired properties. For sources with multiple {\tt DETID}s linked to the same
SRCID\_DR11, we calculated the \ac{S/N} using the source and background counts in
the spectrum and sorted them in descending order. The detection with the
highest \ac{S/N} was selected. The outcome of this initial step was utilised to add
the SRCID\_DR11 associations to each spectrum in C2. Finally, the results of
the last two steps were concatenated, resulting in a total of $210\,444$
sources. Among these, $\dFourOne$ and $\dFourTwo$ are sourced from C1 and C2,
respectively.

Then, we conducted a cross-match between our dataset and \cite{Tranin2022}
using sky coordinates and a matching radius of $1\arcsec$. This was necessary
because the sources in that study were obtained from 4XMM-DR10 and the {\tt
SRCID} do not have a continuity between releases of the catalogue (although
most of them match). The number of detections/sources with extracted/merged
spectra with classifications from \cite{Tranin2022} ultimately amounted to
$92\,238$, with $76\,610$ of them being \acp{AGN}. From this \ac{AGN} subset, we
selected those with \ac{AGN} probability $\geq 95\%$, as calculated by
\cite{Tranin2022}, and with Galactic latitudes $|b|>20^\circ$. This criterion
yielded $35\,538$ \acp{AGN}. Out of these, $8\,467$ had spectroscopic redshifts, and
the rest had photometric redshifts, as explained in
Sect.~\ref{sec:data_photoz}.

The C4 includes $\dFourToFit$ sources in total. The distribution by
classification is as follows: $\dFourToFitAGN$ \acp{AGN}, $\dFourToFitStar$ stars,
$\dFourToFitXRB$ \acp{XRB}, and $\dFourToFitCV$ \acp{CV}. Among these sources,
$\dFourDetThereZero$, $\dFourDetThereOne$, and $\dFourDetThereTwo$ have been
detected exclusively in \pn{}, \mos{}, or both \pn{} and \mos{}, respectively.
Applying the same criteria described in the previous sections, we ended up with
$50\,956$ sources in the \textit{Good sample}. Out of these sources, $41\,142$
meet the requirements and are included in the \textit{Good fit} sample. Of these,
$30\,814$ of them are classified as \acp{AGN}, $9\,353$ as stars, $883$ as \acp{XRB}
and $92$ as \acp{CV}.

In R21, spectral fitting results are included in the $0.5-10\,\mathrm{keV}$ band for
30\,816 source detections, corresponding to 22\,677 unique sources, while C4
includes fits to 35\,538 unique \acp{AGN} in the $0.2-12\,\mathrm{keV}$ band. Compared to
XMMFITCAT, XMMFITCAT-Z used \ac{BXA} for improved sampling of the parameter space,
as we did, but they used the \texttt{wstat} implementation of the Cash
statistic in Sherpa. \texttt{wstat} approximates background modelling by
assigning one free parameter per background bin. This approach can lead to
biased
estimates\footnote{\url{https://giacomov.github.io/Bias-in-profile-poisson-likelihood/}}.
In contrast, we used an empirical background model, fitted to the background
spectrum and then fixed (apart from the normalisation) in the source+background
fit. They fitted two simple and two more complex models to their \acp{AGN}, while we
fitted only one, an intrinsically absorbed power-law, in common with them. We
both used a similar method for the \ac{GoF}. A comparison with their results on the
search for absorbed sources is given in Sect.~\ref{sec:science}.

\subsection{Summary of fitting approaches across the four catalogues}

To aid comparison across the four catalogues, we briefly summarise the key
methodological aspects here. Catalogues C1, C2, and C4 share a common spectral
fitting framework: spectra are binned to a minimum of one count per bin,
background and source+background models are fitted using the Cash statistic
\citep{Cash1979}, and the background is treated via an empirical model whose
parameters (except normalisation) are fixed in the final fit. Catalogue C3
differs in that it uses count rates in five predefined energy bands to
construct low-resolution spectra, which are modeled using $\chi^2$ statistics.

The catalogue C1 includes all 4XMM-DR11 detections with $\geq 100$ net counts,
fitted with an absorbed power-law. C2 focuses on stacked spectra from
4XMM-DR11s sources with multiple detections, using both absorbed power-law and
blackbody models. C4 applies the same fitting methodology as C1 and C2 but uses
classification information to assign appropriate models (e.g. APEC for stars,
bremsstrahlung for \acp{CV}). The background fitting criteria, \ac{GoF} thresholds,
and model priors are consistent across C1, C2, and C4.

\section{%
  Overview of the spectral models, source classification and calculation of
  photometric redshifts%
}
\label{sec:models}

In this section, we describe the models used for the X-ray spectral fitting of
the sources in each one of the four catalogues we complied. We also explain how
we classified the sources included in C4 and calculated their photometric
redshifts.

\subsection{Background model fitting}
\label{sec:back_models}

For the background model used for the fitting of the spectra, we employed the
\XMM{} empirical background model integrated into the {\sc
bxa.sherpa.background} module \citep{Buchner2014}. This model is composed of
two main components: one addressing the cosmic X-ray background and X-ray
emissions from the local hot bubble and Galactic halo, and another component
focused on modelling the camera background, including line contributions.
Importantly, the latter component is not subjected to the instrumental
response.

The background model is a combination of empirical components, including power
laws, Gaussian lines, and thermal emission, inspired by the approach described
in \citet{Maggi2014}. It is fitted through a multi-step process designed to
handle the large number of free parameters and to adapt flexibly to different
camera configurations and background conditions. To verify the robustness of
this model, we performed tests in which the number of free parameters was
reduced by using a simplified background model. These tests were applied to a
representative subsample of sources and showed that the derived source
parameters (e.g., photon index and intrinsic absorption) were consistent with
those obtained using the full background model, with differences well within
the statistical uncertainties and following a one-to-one correlation. These
results confirm that the empirical background model used within \ac{BXA} adequately
captures the background structure without introducing significant biases in the
source spectral parameters.

To evaluate \ac{GoF} for the background spectra, we used a procedure based on
approximating the Cash statistic with a chi-square distribution, since the Cash
statistic does not provide a direct measure of \ac{GoF}. All background spectra
were fitted using the Cash statistic in the Poisson regime with \ac{BXA}, but to
compute $p$-values we followed a two-step procedure. First, we determined the
effective number of free parameters for the \pn{} and \mos{} cameras,
acknowledging that not all background model components were required in every
case. To do this, we used a subset of spectra with at least $1\,000$ background
counts and binned into 30 energy bins over the $0.2-12\,{\rm keV}$ range. We
then compared the resulting C-stat values to chi-square distributions with
varying degrees of freedom to infer the effective number of parameters, finding
values of 12 for \pn{} and 7 for \mos{}.

In the second step, we re-binned the background spectra to have at least 20
counts per bin (reducing to 10 in rare cases of low background), and calculated
$p$-values by comparing the observed $\chi^2$ values (obtained after re-binning)
to the corresponding chi-square distribution using the effective number of
parameters. Although admittedly 10 counts per bin is on the low side,
we point out that almost always the sources in the "Good fit" sample (see
below) have more than 100 counts in both their background and source+background
spectra, effectively comparable to spectra binned at 20 counts, a more usual
binning size. In cases where the number of bins was too small to allow a
statistically meaningful $p$-value (i.e., fewer bins than the effective number
of free parameters), we reduced the number of effective parameters accordingly
to ensure a valid degrees-of-freedom estimate. Detections requiring such
adjustments were flagged and excluded from the “Good fit” sample. A $p$-value
threshold of 0.01 was applied to define the “Good fit” sample. We emphasise
that this procedure was used solely for flagging based on the background fit
quality; all spectral fits themselves were carried out using the Cash
statistic.

For the purpose of assessing the quality of the fit, a $\chi^2$ $p$-value of
$\geq 0.01$ was considered acceptable. In cases where the spectra were
available in both cameras and the $\chi^2$ $p$-value in one camera fell below
this limit while the other exceeded it, only the source spectrum of the latter
camera was taken into consideration.

\subsection{Source spectral models for C1, C2 and C3}
\label{SpecModels_C123}

We employed an absorbed power-law model as the source model for C1, C2 and C3.
To determine the flux within the $0.2-12.0\,{\rm keV}$ range, we incorporated
the {\tt cflux} model in Xspec: ${\tt cflux * tbabs * powerlaw}$. In instances
where the object is observed by both cameras, we introduced an \ac{IIN}
constant defined as \mos{}/\pn{}, using the {\tt const} model in
Xspec. Although an absorbed power-law is the model of choice for \acp{AGN} (see
below), for the limited resolution of CCD X-ray spectra with moderate spectral
quality, it is sufficiently flexible to provide a reasonable result for most
sources. The parameters left unconstrained include the logarithm of the
neutral hydrogen column density of the absorber $N_{\rm H}$, allowed
to vary between 20 and 26 (in log ${\rm cm^{-2}}$), the power-law
photon index $\Gamma$, that varies between 0 and 6, the logarithm of the flux
in the $0.2-12\,{\rm keV}$ band, varying between $-17$ and $-7$ (in log $\rm
erg/cm^2/s$), and the \ac{IIN} constrained between 0 and 5. \ac{BXA} necessitates
the specification of a probability prior for each free parameter in the model,
and we opted for flat priors for all four parameters in the intervals above.

In addition, for C2 we also fitted an absorbed blackbody (${\tt cflux * tbabs *
bbody}$ in Xspec), which could provide a better fit for Galactic sources. With
respect to the parameters and ranges given above for the absorbed power-law,
the only changes are the adjustment of the $N_{\rm H}$ lower limit from 20 to
18, and the replacement of the photon index by the blackbody temperature $kT$,
allowed to vary between $0.01-10\,{\rm keV}$, also with a flat prior.

Since we have used Cash statistics for the fits and a large fraction of the
spectra have less than 100 net counts, we have decided not to use $\chi^2$ as a
\ac{GoF} indicator. The Cash maximum likelihood statistics lacks a direct estimate
of \ac{GoF}\@. Therefore, we used the method proposed by \cite{Buchner2014}, also
followed in R21. We calculated the Kolmogorov-Smirnov (KS) statistic between
the observed and expected data+model counts, and the corresponding $p$-value,
as a quantitative estimate of the \ac{GoF}\@. We note however that in this case the
$p$-values for the KS statistic cannot be calculated the usual way. The
cumulative distribution of the model depends on the parameters that were
estimated from the data distribution. This implies that the two compared
distributions are not independent. Nevertheless, we can do a permutation test
to get an estimate of the $p$-value. For each source, we did $1\,000$
resamplings, rearranging the original data+model sample in two equal-size
subsamples, where the counts in each energy bin can come either from the data
or the model sample, and estimate the corresponding KS statistic. Our estimated
$p$-values are the fraction of resamplings that have statistics larger than the
statistic of the original samples. Any model showing a KS $p \geq 0.01$ is
considered as an acceptable fit.

While several studies
\citep[e.g.,][]{%
  Buchner2014,%
  Marchesi2016,%
  Masini2020,%
  Liu2022,%
  Peca2023,%
  Boorman2025%
}
have highlighted the importance of including a soft X-ray component in \ac{AGN}
spectral models, such as scattered power-law emission or thermal excess, our
present work adopts a simpler approach using a single absorbed power-law. This
choice was motivated by the limited photon statistics in many of our sources,
the need to maintain a manageable number of free parameters in automated fits,
and a uniform approach over the full catalogue. We acknowledge that the lack of
a soft component can introduce biases in the estimation of spectral parameters,
particularly for obscured sources. A future extension of this work will explore
multi-component models in a subset of well-exposed \acp{AGN} to quantify this effect.

We note that the spectral model used for C3 contain at most four free
parameters: normalisation, photon index, absorption, and the \ac{IIN} (the latter
parameter is free only if spectra from both types of EPIC camera are used in
the fit). Given that the spectra consist of five coarse flux bins across XMM
bands, this ensures that the number of free parameters does not exceed the
number of data points ($N=5$), preserving statistical robustness.

For all catalogues the mode and median values of each calculated parameter are
provided, along with the narrowest interval that includes 90\% of the
probability, and percentiles of 5 and 95 per cent. Throughout this paper, we
use the mode values of the presented parameters.

\subsection{Source classification}
\label{sec:data_class}

To categorise the X-ray sources included in C4, a probabilistic technique using
a naive Bayes classifier was devised, which is thoroughly explained in
\cite{Tranin2022}. In essence, this approach drew inspiration from its
intuitive characteristics, extending from the basic classification principles
seen in rudimentary decision trees. To carry out the classification of X-ray
sources, specific data columns from the \XMM{} catalogue's 4XMM-DR10 version
were employed.

The catalogue was also expanded with multi-wavelength counterparts, employing
the NWAY algorithm \citep{Salvato2018} and using a number of available
catalogues (e.g., Gaia, GLADE) as described in \cite{Tranin2022}. A dataset of
25\,160 previously identified sources was generated and categorised into
distinct subgroups, encompassing \acp{AGN}, stars, \acp{XRB}, and \acp{CV}. The
probability density for various properties associated with each source type was
estimated and these probabilities were utilised to assess the likelihood of
classifying the sources. In cases where a property value was missing, the
likelihood was substituted with the probability that a source of that class
would have a missing value for that property. Subsequently, Bayes' rule was
applied, considering each property for each source type, to determine the
probability associated with the source's nature. Additionally, an outlier class
was introduced to identify rare sources of other types. The algorithm's
performance was validated using a test sample, yielding outstanding precision
results: 97.2\% for \acp{AGN}, 98.9\% for stars, 93.7\% for \acp{XRB}, and
84.6\% for \acp{CV} \citep{Tranin2022}.

\subsection{Photometric redshifts}
\label{sec:data_photoz}

For the X-ray sources classified as \acp{AGN}, based on the source classification
method outlined in the previous sub-section, we performed the computation of
photometric redshifts. To achieve this, we employed the methodology detailed in
\cite{Ruiz2018}. In a nutshell, this approach leverages optical counterparts
from datasets like SDSS or PanSTARRS for the X-ray sources and, whenever
feasible, also seeks counterparts in the near-infrared (e.g., 2MASS, UKIDSS,
VISTA-VHS) and/or mid-infrared bands (AllWISE).

To facilitate the cross-correlation of multiple catalogues, we utilised the
`xmatch' tool from the astromatch
package\footnote{\url{https://github.com/ruizca/astromatch}}. This tool
facilitated the matching of multiple catalogues and provided Bayesian
probabilities for associations or non-associations, as detailed in
\cite{Pineau2017, Ruiz2018}. Subsequently, a machine-learning (ML) technique
was applied to compute photometric redshifts. Specifically, we employed the
MLZ-TPZ method, as described in \cite{Kind2013}, which relies on a supervised
technique involving prediction trees and random forest. MLZ-TPZ is a Python
package that can be executed in parallel, enabling the swift and reliable
calculation of photometric redshifts along with their corresponding probability
density functions (PDF).

The cross-matches and the photometric redshifts were based on preliminary
results from \xmmiiathena{} at the time this work was done, and are available
from the authors upon request. The final catalogues are available in the
project web pages and will be fully presented in Nebot et al.\ (in preparation)
and Ruiz et al.\ (in preparation), respectively.

This preliminary catalogue of photometric redshift was built using the training
sample presented by \citet{Mountrichas2017a} and \citet{Ruiz2018}; it contains
$\sim5000$ X-ray selected \acp{AGN} with optical counterparts in SDSS or PanSTARRS
and reliable spectroscopic redshifts. More than 90 per cent of the training
sources have additional photometry in the near- and/or mid-IR\@.

The statistical accuracy and reliability of our photometric redshift was
estimated through the widely used normalised median absolute deviation
$\sigma_{\rm NMAD}$ and the percentage of catastrophic outliers
$\eta$.\footnote{See e.g.\ Eqs.~1-4 from \citet{Ruiz2018} for the formal
definition of these statistical indicators} For sources with SDSS (PanSTARRS)
photometry, $\eta$ ranges from 9 (4) per cent for extended sources with
additional photometry in the near- and mid-IR to 29 (41) per cent for
point-like sources (i.e.\ optical emission dominated by the \ac{AGN}) with
photometric information only in the optical bands. Changes in $\sigma_{\rm
NMAD}$ are less significant depending on the optical morphology and the amount
of photometric information, ranging from $\sim0.08$ in the worst case (only
optical photometry) to $\sim0.04$ in the best case (full photometry in the
optical and IR bands). For a detailed analysis of this cross-validation and
potential explanations for the differences between SDSS and PanSTARRS results,
see \citet{Ruiz2018}. We note that in this work, we use only the best-fit point
estimates of the photometric redshifts in the spectral fitting.

\subsection{Source spectral models for C4}
\label{SpecModels}

For sources in the C4 catalogue, different models were used, based on the
classification of the sources, as described in the following sections. The
number of sources in each category, including the \textit{Good sample} and
\textit{Good fit} sample are listed previously in Sec.~\ref{sec:C4_analysis}

\subsubsection{AGNs}

For \acp{AGN}, a redshifted absorbed power-law model with Galactic absorption
was utilised, the local Galactic absorption with $N_{\rm H}$ fixed to the total
$N_{\rm H}$ in that line of sight, plus in-situ absorption at the \ac{AGN}
redshift with free $N_{\rm H}$. Specifically, the model employed is ${\tt cflux
* tbabs(ztbabs * zpowerlaw)}$ and we utilise the same priors for the parameters
as in C1 (Sec.~\ref{SpecModels_C123}). The power-law is thought to arise from
upscattering of photons from a hot corona, while the absorption is associated
with an obscuring circumnuclear structure (the `torus'). The redshift used is
either the spectroscopic redshift (when available) or the mode of the
photometric redshift PDF and it is always kept fixed at the $z_{\rm best}$
value in the catalogue file. About $8\,500$ redshifts are spectroscopic.

\subsubsection{Stars}

For stars we fit a single APEC (Astrophysical Plasma Emission Code) plasma
\citep{Dere1997, Smith2001}. APEC models the X-ray emission arising from the
stellar corona. The corona is a region of highly ionised gas surrounding a
star, characterised by high temperatures (typically millions of degrees
Kelvin). This hot plasma emits X-rays which are modelled by the APEC model.
APEC incorporates various physical parameters of the plasma, such as
temperature, elemental abundances, and emission measure, to calculate the
expected X-ray spectrum. In Xspec notation the model reads ${\tt cflux * tbabs
* apec}$. We utilise the same priors as in the C2 absorbed blackbody model
(Sec.~\ref{SpecModels_C123}), but modify slightly the upper limit of the APEC
plasma temperature $kT$ to vary between 0.01 and 17 keV.

\subsubsection{X-ray binaries}

The \acp{XRB} are binary star systems in which the accretion process
from a compact object, such as a neutron star or a black hole, from their
companion normal star, can produce intense X-ray emission. The X-ray spectra of
\acp{XRB} are typically modeled using a blackbody and a power-law component.

In the case of neutron stars, the blackbody component in \ac{XRB} spectra
represents thermal emission from a hot surface, with the temperature of the
blackbody related to the surface temperature of the neutron star. For black
holes, which have no surface, the thermal emission is often modeled using a
disk blackbody component to represent the accretion disk's emission. It is
important to note that the blackbody results differ from the typical disk
blackbody ones.

The power-law component in \ac{XRB} spectra is associated with non-thermal
processes, often related to the corona surrounding the compact object. The
power-law index characterises the shape of the non-thermal emission spectrum.
High-energy processes, such as inverse Compton scattering, can contribute to
the power-law component. While the accretion disk emits thermal emission, it
also reflects Comptonised photons, which are not directly related to the disk
itself.

In Xspec notation the combined blackbody and power-law model reads ${\tt
cflux_{bb} * tbabs * bbody + cflux_{pl} * tbabs * powerlaw}$, where the two
components of the flux refer to the two separate model components, and the $\tt
tbabs$ is the same between the two flux components. We adopt the same priors
for the parameters as in C1 and C2 (Sec.~\ref{SpecModels_C123}), while allowing
$N_{\rm H}$ to vary between 18 and 24 (in log cm$^{-2}$)

\subsubsection{Cataclysmic variables}

The \acp{CV} are binary star systems consisting of a white dwarf
primary star and an usually main-sequence secondary star. The white dwarf
accretes matter from its companion, leading to various types of outbursts and
transient phenomena. \acp{CV} are characterised by their erratic behavior and can
undergo episodes of increased brightness, such as dwarf novae outbursts or
classical novae explosions.

X-ray emission from \acp{CV} is often associated with the accretion process. As
matter from the secondary star accretes onto the white dwarf, it forms an
accretion disk, and the release of gravitational potential energy results in
high-temperature regions that emit X-rays. One common process responsible for
X-ray emission in \acp{CV} is bremsstrahlung. Bremsstrahlung occurs when
charged particles, such as electrons, are deflected by the strong electric
fields in the vicinity of other charged particles, causing them to emit
radiation. In the context of \acp{CV}, the hot plasma in the accretion disk
emits X-rays through bremsstrahlung. In Xspec notation the bremsstrahlung model
reads ${\tt cflux * tbabs * bremss}$. We utilise the same priors as in the C2
blackbody model (Sec.~\ref{SpecModels_C123}), but allow the bremsstrahlung
plasma temperature $kT$ to vary between 0.0001 and 200 keV.

\begin{figure*}[htbp]
  \centering
  \includegraphics[width=0.9\linewidth, height=13cm]{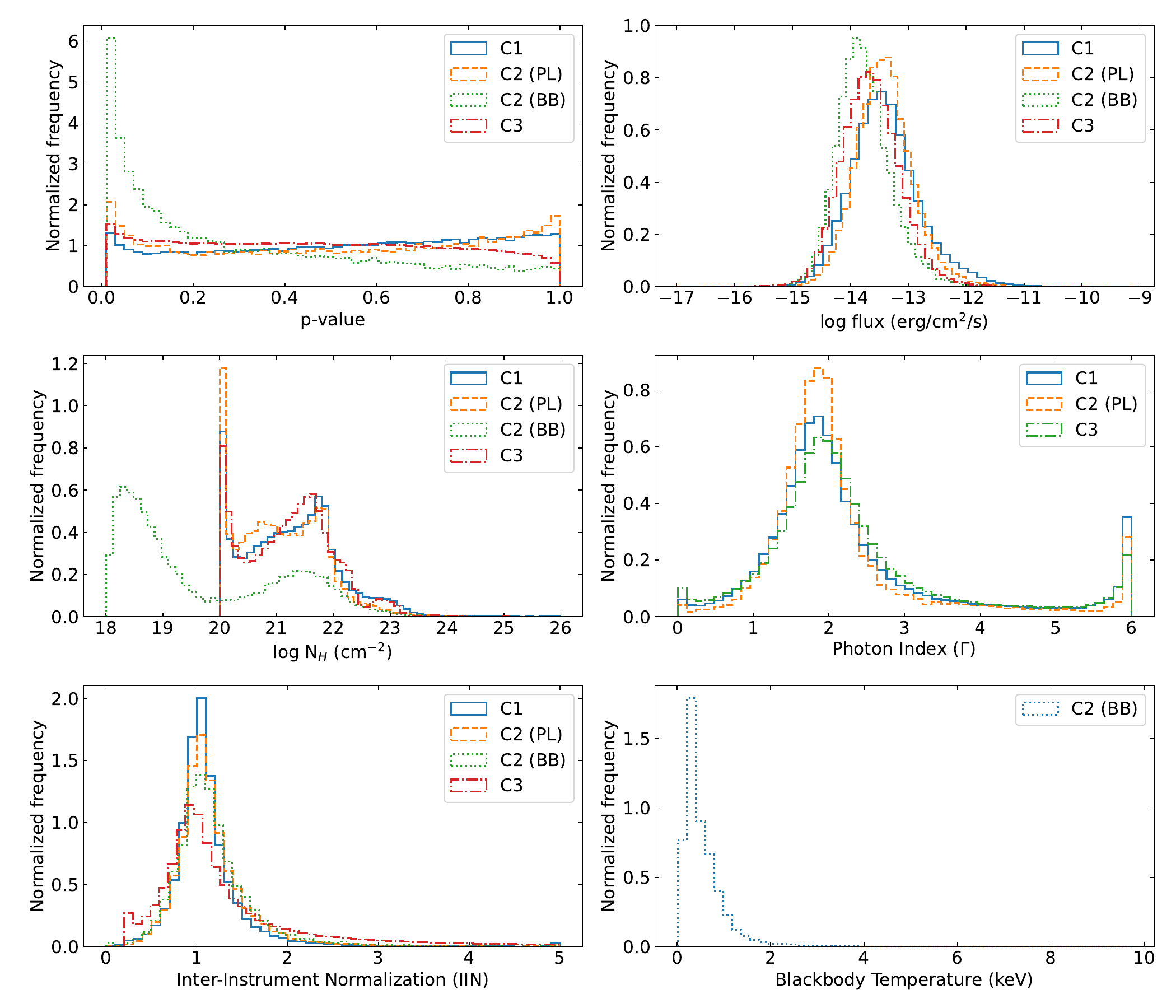}
  \caption{%
    Distributions of the $p$-values (top-left panel), flux (top-right panel),
    $N_{\rm H}$ (middle-left panel), photon index (middle-right panel), \ac{IIN}
    (bottom-left panel) and blackbody temperature (bottom-right panel) of the
    sources included in the \textit{Good fit} samples of C1, C2 and C3, as
    indicated in the legends.
  }
  \label{fig_merged_distrib}
\end{figure*}

\begin{table*}
  \centering
  \caption{%
    Median values of (the mode of) each parameter for each source in the
    catalogues C1-4.
  }
  \begin{tabular}{lrrrrr}

  \toprule
    catalogue &
    $\mylogfx{}$ &
    $\mylognh{}$ &
    $\Gamma$ &
    $kT$ (keV) &
    ${\rm IIN}$ \\

  \midrule
  C1         & ${-13.44}_{ -0.52}^{ +0.60}$ & ${ 21.26}_{ -0.98}^{ +0.73}$ & ${  1.95}_{ -0.58}^{ +1.34}$ &                           -- & ${  1.05}_{ -0.21}^{ +0.29}$ \\

  \midrule
  C2 (pl)    & ${-13.43}_{ -0.45}^{ +0.47}$ & ${ 21.04}_{ -0.86}^{ +0.79}$ & ${  1.91}_{ -0.47}^{ +0.86}$ &                           -- & ${  1.08}_{ -0.24}^{ +0.37}$ \\
  C2 (bb)    & ${-13.80}_{ -0.40}^{ +0.47}$ & ${ 19.06}_{ -0.72}^{ +2.46}$ &                           -- & ${  0.40}_{ -0.19}^{ +0.47}$ & ${  1.10}_{ -0.28}^{ +0.41}$ \\

  \midrule
  C3         & ${-13.74}_{ -0.60}^{ +0.54}$ & ${ 21.28}_{ -1.00}^{ +0.68}$ & ${  2.03}_{ -0.66}^{ +1.20}$ &                           -- & ${  1.08}_{ -0.36}^{ +1.08}$ \\

  \midrule
  C4 (\acp{AGN}) & ${-13.34}_{ -0.41}^{ +0.44}$ & ${ 20.73}_{ -0.64}^{ +1.15}$ & ${  1.96}_{ -0.42}^{ +0.45}$ &                           -- & ${  1.05}_{ -0.19}^{ +0.26}$ \\
  C4 (stars)     & ${-13.88}_{ -0.45}^{ +0.49}$ & ${ 20.28}_{ -1.83}^{ +1.60}$ &                           -- & ${  0.67}_{ -0.38}^{ +0.56}$ & ${  1.06}_{ -0.20}^{ +0.30}$ \\
  C4 (XRBs)      & ${-13.51}_{ -0.45}^{ +0.60}$ & ${ 21.32}_{ -0.64}^{ +0.57}$ & ${  1.73}_{ -0.66}^{ +0.69}$ & ${  0.06}_{ -0.05}^{ +0.62}$ & ${  1.03}_{ -0.26}^{ +0.30}$ \\
  C4 (CVs)       & ${-12.91}_{ -0.45}^{ +0.84}$ & ${ 21.00}_{ -1.02}^{ +0.57}$ &                           -- & ${  4.65}_{ -3.45}^{+42.49}$ & ${  1.05}_{ -0.08}^{ +0.17}$ \\

  \bottomrule
  \end{tabular}
  \tablefoot{%
    The columns correspond to the flux, hydrogen column density, power-law (pl)
    photon index, \ac{IIN}, blackbody (bb) or bremsstrahlung temperature,
    respectively. The errors correspond to the 16 and 84 percentiles.
  }
  \label{tab:table_properties_c1234}
\end{table*}

\section{Results}
\label{sec:results}

In this section, we describe the main properties calculated by following the
analysis we applied on the four catalogues, as described in
Sect.~\ref{sec:catalogues_analysis}. In all cases, the measurements from the
\textit{Good fit} subsets of each catalogue is presented.

\subsection{Main properties of sources included in C1}
\label{sec:properties_C1}

As previously mentioned, the C1 catalogue includes the results from fitting an
absorbed power law model to all extracted spectra of detections in the
4XMM-DR11 dataset (Sect.~\ref{sec:C1_analysis}). The top-left panel of
Fig.~\ref{fig_merged_distrib} presents the distribution of the $p$-values of
the fits for the sources included in the \textit{Good fit} subset of C1, as
indicated in the legend. The top-right panel presents the distribution of the
fluxes of the sources. As previously mentioned, fluxes are obtained in the 0.2
to 12.0 keV band, by including the cflux component in our model. In case of
multiple camera spectra for a detection, the reported flux is the \pn{} flux.
The median value of the mode calculations of the flux is $\mylogfx = -13.44$,
with a scatter (standard deviation) of 0.6
(Table~\ref{tab:table_properties_c1234}). The scatter is estimated by first
computing the differences between individual flux values and the median flux.
The standard deviation of these differences is then calculated to quantify the
typical dispersion of values around the median, providing a measure of
statistical scatter.

The middle left and right panels of Fig.~\ref{fig_merged_distrib} present the
measurements for the spectral parameters, namely the hydrogen column density
$N_{\rm H}$ and photon index $\Gamma$, respectively. The median values (of the
mode calculations) are $\mylognh = 21.26$ and $\Gamma = 1.95$. We notice a
second lower tail in the $\Gamma$ distribution at very high values ($\approx
6$). This is probably caused by trying to fit thermal emission with an absorbed
power law. There is also a peak at $\mylognh \sim 20$. We note that these
extreme values close to the chosen limits of the priors should be treated as
lower or upper limits of their respective parameters (see
Sec.~\ref{sec:models}). A cross-match of our dataset with the SIMBAD database
\citep{Wenger2000}, reveals that this peak is mainly populated by stars.

The bottom-left panel of Fig.~\ref{fig_merged_distrib} presents the \ac{IIN},
defined as \mos{} over \pn{}. The median value of the (mode) \ac{IIN} is
$1.05$. Previous measurements of the \XMM{} \ac{IIN}\@ range from $1.02-1.08$
based on 2XMM \citep{read2014A&A...564A..75R}, and $\sim 1.04-1.17$ for 3C 273
and PKS 2155-304 \citep{madsen2017AJ....153....2M}. Our values are consistent
with these measurements especially so given the broad distribution in
\ac{IIN}\@. About 1\% of the sources (4\,052) present an \ac{IIN} above two or
below 0.5. This can be explained by the fact, that if two \mos{} cameras are
present then their spectra were combined. This merging included cases where the
combined spectra are taken in two different camera submodes. In addition, the
submodes between the \pn{} and \mos{} cameras may also differ. We checked the
relative contribution to the overall and the extreme cases for the different
camera submode combinations. A significantly enhanced contribution to the
extreme cases is only found for observations where any \pn{} submode is
combined with the `Fast (Un)compressed' mode of one or both \mos{}. However,
this cannot explain the number of sources we get with very high or low \ac{IIN}
values, as there are only a few detections with this combination of submodes
(0.7 per cent of the detections in the \textit{Good fit} sample that have \pn{}
and \mos{} spectra).

These results may be directly compared to the Chandra Source Catalog (CSC) 2.1
\citep{Evans2020}, where a similar approach of fitting an absorbed power-law to
detected sources is employed. Their master source table contains $407\,806$
unique sources from $15\,533$ Chandra observations. The total area is
$730\,{\rm deg^2}$, which decreases to $705\,{\rm deg^2}$ and $137\,{\rm
deg^2}$ at fluxes fainter than $<10^{-13}$ and $<10^{-15}$ (in any Chandra
band), respectively. We select $86\,368$ sources at $>5\sigma$ flux
significance and culling flagged sources. The corresponding CSC 2.1 power-law
median, 16th, and 84th percentiles are $\Gamma = 2.03_{-0.41}^{+1.17}$ and
$\mylognh = 21.44_{-0.80}^{+0.72}$, which are in agreement with our reported
values within the errors.

\subsection{Main properties of sources included in C2}
\label{sec:properties_C2}

The C2 catalogue, includes the results of fitting an absorbed power-law model
and an absorbed blackbody model to the merged spectra of all sources of the
4XMM-DR11s catalogue with more than one contributing observation (see
Sect.~\ref{sec:C2_analysis}). Fig.~\ref{fig_merged_distrib} displays the
distributions of the $p$-values, $f_{\rm X}$, $N_{\rm H}$, \ac{IIN}, $\Gamma$ and
the blackbody temperature of the two \textit{Good fit} samples, as indicated in
the legend. The median values of the mode calculations, for the power-law and
the blackbody models, respectively, are: $\mylogfx = -13.43$ and $-13.80$,
$\mylognh = 21.04$ and $19.06$, ${\rm IIN}=1.08$ and $1.10$, respectively
(Table~\ref{tab:table_properties_c1234}). The median values of the photon
index, calculated for the power-law model is $1.91$ and the median blackbody
temperature is $0.40\,{\rm keV}$. About the blackbody temperature, we note that
there are some sources with extreme $kT$ values up to $9.7\,{\rm keV}$\@.
However, less than one per cent (88 sources) have $kT$ values higher than
$3\,{\rm keV}$\@.

About five per cent of the sources have \ac{IIN} above 2 or below 0.5, independent
of the sources model. This fraction is lower with increased count rate, and is
stronger for the blackbody than for the power-law model. We also note that this
fraction is higher in C2 compared to C1. For C2, we do not only combine
spectra of the two \mos{} cameras and fit spectra of different \pn{} and \mos{}
submodes simultaneously, but we also combine spectra of different observations.
This can be a reason for the higher percentage of extreme cases and the reduced
dependence on count rate.

Regarding the spectral parameters, the distribution of the photon index
parameter calculated for the power-law model, similarly to the C1 catalogue,
presents a second, lower peak at high $\Gamma$ values. As mentioned earlier,
this could be due to trying to fit thermal emission with an absorbed power-law.

A comparison between the fluxes and \ac{IIN} values calculated using the
power-law and blackbody models reveals strong agreement between the two
approaches. However, the power-law model generally produces higher flux
estimates than the blackbody model. This difference likely arises because most
X-ray sources are expected to be \acp{AGN}, for which an absorbed power-law typically
provides a more accurate representation than a blackbody (see
Appendix~\ref{appendix_c2}).

\begin{figure*}[htbp]
  \centering
  \includegraphics[width=0.90\linewidth, height=14cm]{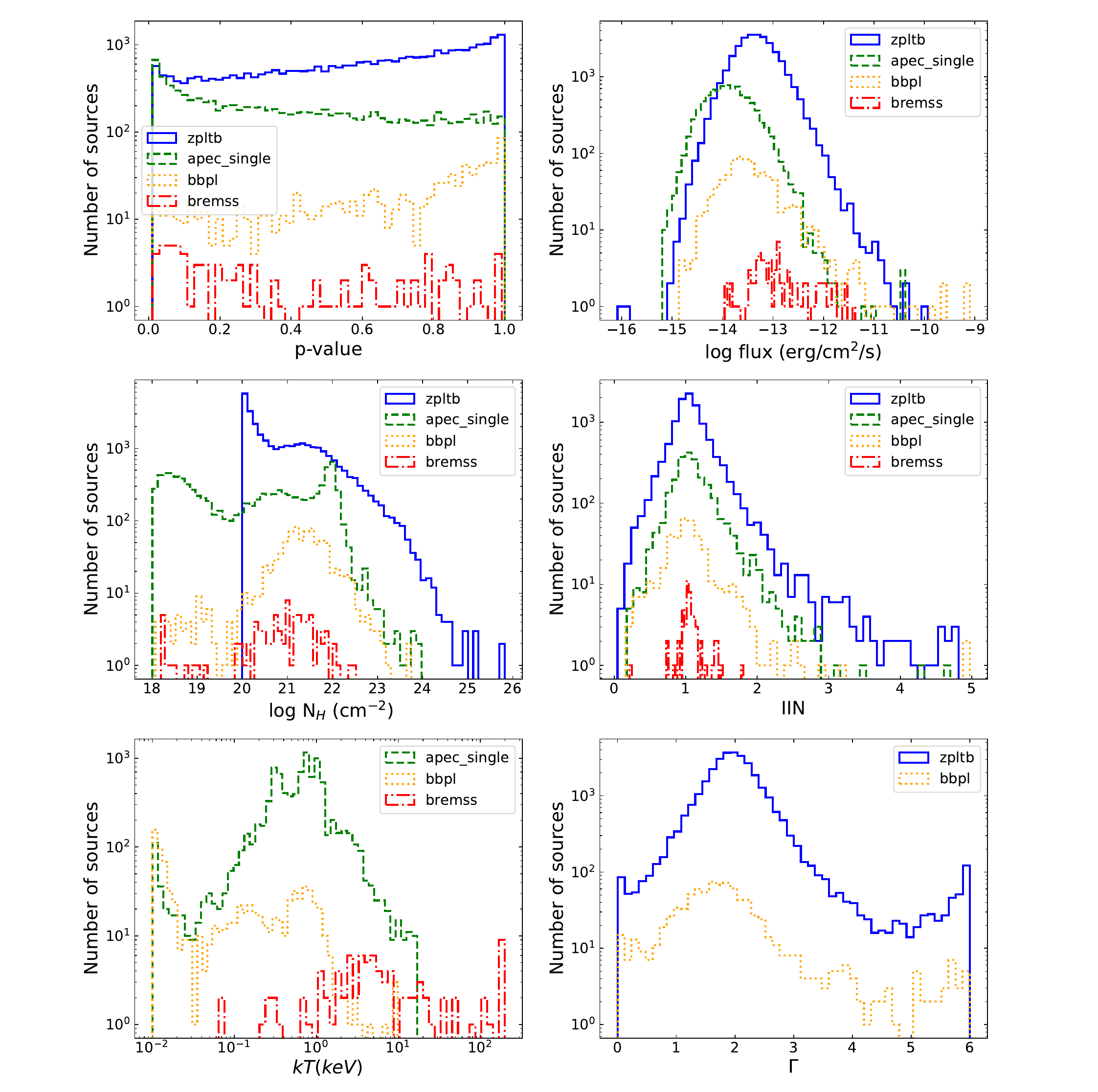}
  \caption{%
    Distributions of the $p$-values (top-left panel), flux (top-right panel),
    $N_{\rm H}$ (middle-left panel), \ac{IIN} (middle-right panel), blackbody
    temperature (bottom-left panel) and photon index (bottom-right panel) of
    the sources included in the \textit{Good fit} sample of C4. Blue lines
    present the results for \acp{AGN} (\zpltb{} model), green lines display the
    measurements for stars (\apec{} model), orange lines show the calculations
    for \acp{XRB} (\bbpl{} model) and red lines illustrate the results for
    \acp{CV}
    (\bremss{} model).%
  }
  \label{fig:C4_distrib}
\end{figure*}

\subsection{Main properties of the detections included in C3}
\label{sec:properties_C3}

As mentioned earlier, the C3 catalogue includes the results of the fitting of
the spectra, constructed using the count rates in five energy bands, from the
4XMM-DR11 catalogue. Fig.~\ref{fig_merged_distrib} presents the distributions
of the various parameters calculated by fitting the \textit{Good fit} subset of
C3, as indicated in the legend. The median value of $\mylogfx = -13.74$
(Table~\ref{tab:table_properties_c1234}). The median value of the neutral
hydrogen column density is $\mylognh = 21.28$ and the median value of photon
index is $\Gamma=2.03$ that is close to the expected value for a population
dominated by \acp{AGN} \citep[$\approx 2$;][]{Nandra1994}, and similar to the
$\Gamma$ values obtained for the C1 and C2 catalogues
(Table~\ref{tab:table_properties_c1234}). There is also a non-negligible
number of sources in the extremes of our selected photon-index interval. The
spectral model we used is probably not adequate for these ultra-hard/ultra-soft
sources. The median values of \ac{IIN} is $1.08$, suggesting the possible
inaccuracies in the cross-calibration of \XMM{} cameras are small.

In the subsections about catalogues C1 and C2, the merging of \mos{} spectra and
the combination of observations were mentioned as reasons why extreme \ac{IIN}
values are obtained. However, none of these cases apply to C3. It was also
stated that combining different submodes of \pn{} and \mos{} can only explain a small
fraction of those cases. Therefore, previous sections do not fully account for
why extreme values of \ac{IIN} are observed in C3. In C3, the extreme \ac{IIN} values are
more likely due to the limitations and simplifications of the spectral models
used, as well as potential issues with the data quality or the presence of
peculiar sources that are not well-represented by the applied models.

To evaluate the differences between using count rate spectra (C3) and applying
proper spectral fitting (C1), we compare the calculated values for $f_{\rm
X}$, $N_{\rm H}$, $\Gamma$, and \ac{IIN} between the C3 and C1 catalogues (see
Appendix~\ref{appendix_c1_c3}). Our findings indicate that $f_{\rm X}$ values
from the two methods are in good agreement, with mean and median differences of
0.05 and 0.02, and a scatter of 0.24. Correlations for $N_{\rm H}$ and $\Gamma$
are also reasonable but with larger scatter, mainly due to poorly constrained
posteriors in low-count sources. \ac{IIN} values generally cluster around one,
though the correlation is weaker, especially in the C1 catalogue, where \mos{}
spectra were merged. These results support the use of count rate spectra for
population studies, provided proper filtering is applied, while detailed
spectral fitting remains preferable for individual sources.

\subsection{Main properties of sources included in C4}
\label{sec:properties_C4}

The C4 catalogue, as mentioned in Sect.~\ref{sec:C4_analysis}, includes
spectral fitting for sources identified as \acp{AGN}, \acp{XRB}, \acp{CV} and
stars. Fig.~\ref{fig:C4_distrib}, presents the distributions of the
$p$-values, $f_{\rm X}$, $N_{\rm H}$, \ac{IIN}, $kT$ and $\Gamma$ calculated by
fitting the \textit{best-fit} subsets of C4 with the four models corresponding to
their respective classifications. The median values of the mode of each
parameter are shown in Table~\ref{tab:table_properties_c1234}. For the sources
identified as \acp{AGN}, the intrinsic (i.e.\ absorption corrected) rest-frame
$2-10\,{\rm keV}$ luminosity $L_{\rm X}$ has been computed using the chains of
flux and $\Gamma$ measurements obtained from the X-ray spectral fitting
process, and we provide the mode and the narrowest interval that encompasses
90\% of the values. The \acp{AGN} in our dataset have a median $\myloglx \sim
44$.

We find overall strong agreement in $f_{\rm X}$, \ac{IIN}, and $\Gamma$ measurements
between catalogues, with mean and median differences close to zero and moderate
scatter. Comparisons between C4 and C1/C2 confirm the consistency of spectral
parameters, especially for \acp{AGN}\@. For \acp{XRB}, $f_{\rm X}$ values are
systematically higher in C4 due to model differences, while \ac{IIN} shows good
agreement. A significant correlation between flux and $p$-value differences
supports the improved fit of the more complex model in C4. For more details see
Appendix~\ref{appendix_c4_vs_all}.

\section{Science application}
\label{sec:science}

We show in this section one of the potential applications of these catalogues.
We use C4 to assess the optical/MIR colour \ac{AGN} selection techniques, in
particular their capabilities to select X-ray selected absorbed \acp{AGN}, along the
lines of R21. With respect to that work, we have a larger number of unique
sources and we discuss the effects of different thresholds and definitions of
absorbed and unabsorbed sources.

This section is only intended to showcase the scientific potential of the
catalogues. We have not attempted to address the multiple selection effects in
the C4 catalogue. Sensitivity limits for each band and observation/stack are
available in the XSA/stacked web pages, but quantifying the consequences of
using only classified sources with extracted spectra, and selecting just
sources with photometry in all bands and with "good" photometric redshifts, is
beyond the scope of this paper.

In this section we start from the C4 \textit{Good fit} sample. A further
selection in the sample was to use only sources with multi-wavelength
counterparts, extracted during the work for finding multi-wavelength
counterparts and identifications for \XMM{} sources within the \xmmiiathena{}
project. The number of sources at this stage includes $30\,610$ \acp{AGN}, $1\,525$
stars, $50$ \acp{XRB} and $35$ \acp{CV}. The latter two types of sources are shown in
some figures, but we have taken no further interest in them, due to the limited
statistics.

We have matched the stars to the GAIA DR3 catalogue \citep{GAIA2016,GAIA2023}
within 1~arcsec of the sky positions of the multi-wavelength counterparts,
using Vizier and the CDS X-match service, getting matches for $1\,496$ stars.
One of the parameters from GAIA is the effective temperature of the stars
$T_{\rm eff}$. We have defined as \mylowt{} and \myhight{} stars those with
$T_{\rm eff}$ below and above $4\,000$~K, respectively. There are 355 \mylowt{}
stars and 807 \myhight{} stars. That information is missing for 334 stars.

An X-ray luminosity limit commonly used in the literature to select \acp{AGN} is
$10^{42}\,\mathrm{erg\,s^{-1}}$ in the $2-10\,{\rm keV}$ band, since there are
no local pure star-forming galaxies with a luminosity in that range above that
limit. One of the most X-ray luminous galaxies in that category is NGC3256,
with an X-ray luminosity of only $2.5\times10^{41}\,\mathrm{erg\,s^{-1}}$ but
with no evidence for an \ac{AGN} \citep{Moran1999}. Following R21, we have selected
the \acp{AGN} for which the upper 90\% uncertainty limit in their luminosity is
higher than $10^{42}\,\mathrm{erg\,s^{-1}}$. Changing this criterion to be that
the mode of the luminosity being above that limit leads to a reduction of the
samples by $\sim$0.5-1\% and does not change any of the conclusions below.
Furthermore, $<$4\% of the \acp{AGN} selected with our criterion have the lower 90\%
confidence limit on their luminosities below $10^{42}$~erg/s, ensuring that the
vast majority of samples have luminosities above that limit with 95\%
confidence. Finally, we have restricted the sample to \acp{AGN} with redshifts
$z<3.5$, to facilitate the comparison with that R21. There are $29\,935$
\acp{AGN} fulfilling these conditions. In the rest of this section these will
be called the \textit{Good fit} \acp{AGN}.

R21 found a number of sources with X-ray luminosity above $10^{48}$~erg/s,
which they argued were unphysical. They therefore defined a `reliable' sample
(a subsample of the \textit{Good fit} \acp{AGN}) by requiring that the photometric
redshift probability was concentrated in a single peak (parameter
PDF\_PS$\geq$0.7) and that the 90\% confidence interval in the column density
was $\leq$2~dex for `absorbed' sources (see below). We have followed that
definition for ease of comparison with them, but we note here that there are no
sources among the \textit{Good fit} \acp{AGN} with luminosity above that value.
Finally, for their reliable sample, R21 excluded absorbed \acp{AGN} (see below for
definition) with luminosities above $10^{45.3}\,{\rm erg/s}$. Since there are
no high luminosity sources in our sample, we have not introduced any upper cut
on the luminosity to define the reliable sample. We will use this definition of
reliable samples in the rest of this section. The concrete numbers of \acp{AGN}
in them are given below for several definitions of absorbed and unabsorbed
\acp{AGN}.

Having selected a reliable sample, for our main results we will define absorbed
\acp{AGN} (\myabsagn{}) as those with very flat photon indices (upper limit of the
90\% confidence interval on the photon index $\Gamma$ below 1.4) or with large
column densities (lower limit of the 90\% confidence interval on the column
density above $10^{22}\,{\rm cm^{-2}}$) and unabsorbed \acp{AGN} (\myunabsagn{}) all
the rest, as in R21. The former condition is commonly used to select X-ray
absorbed samples \citep[e.g.][]{Corral2014}, and it is based on the observation
that Compton-thick sources show flat X-ray spectra when fitted with a simple
power-law \citep[e.g.][]{Matt1996}. We finally have $29\,935$ \textit{Good fit}
\acp{AGN}, $1\,526$ of which are absorbed and $28\,409$ are unabsorbed. The reliable
sample includes $21\,999$ unique \acp{AGN}, of which $1\,137$ are absorbed and
$20\,862$ are unabsorbed (to be compared with $977$ and $17\,158$ reliable
absorbed and unabsorbed detections in R21). We will examine the effects of this
definition on our conclusions in two ways: first by changing the limit for
absorbed sources to $\mylognh > 23$ (\mylognhxxiii{} limits: $655/29\,280$
\myabsagn{}/\myunabsagn{}, $444/21\,567$ reliable \myabsagn{}/\myunabsagn{}),
and second keeping our default limit on the column density, but defining
unabsorbed sources as those with the upper limit of the 90\% confidence
interval below that limit, the rest being undetermined \acp{AGN} (\myundetagn{})
(\myabsunabsundet{} limits: $1\,526/17\,601/10\,808$
\myabsagn{}/\myunabsagn{}/\myundetagn{}, $1\,137/13\,403/7\,459$ reliable
\myabsagn{}/\myunabsagn{}/\myundetagn{}). This last definition of unabsorbed
\acp{AGN} is more stringent, since for them we can say, with ${\sim} 95$\%
confidence, that the absorption is below our default limit, so they are
genuinely `unabsorbed' (with the definition above, we just cannot tell apart
these objects from e.g.\ those with absorption but large error bars).

\begin{figure}[htbp]
  \centering
  \includegraphics[width=\linewidth, height=4.5cm]{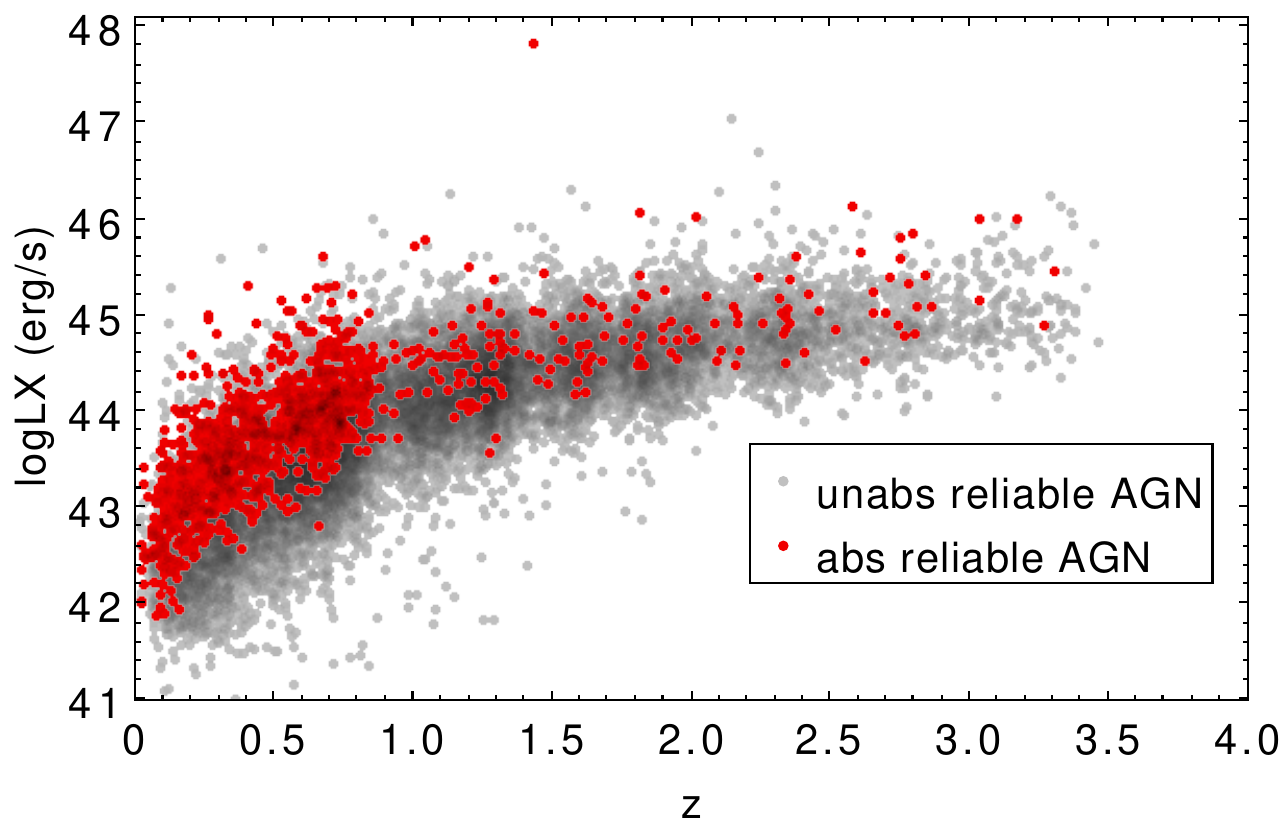}
  \includegraphics[width=0.49\linewidth, height=4cm]{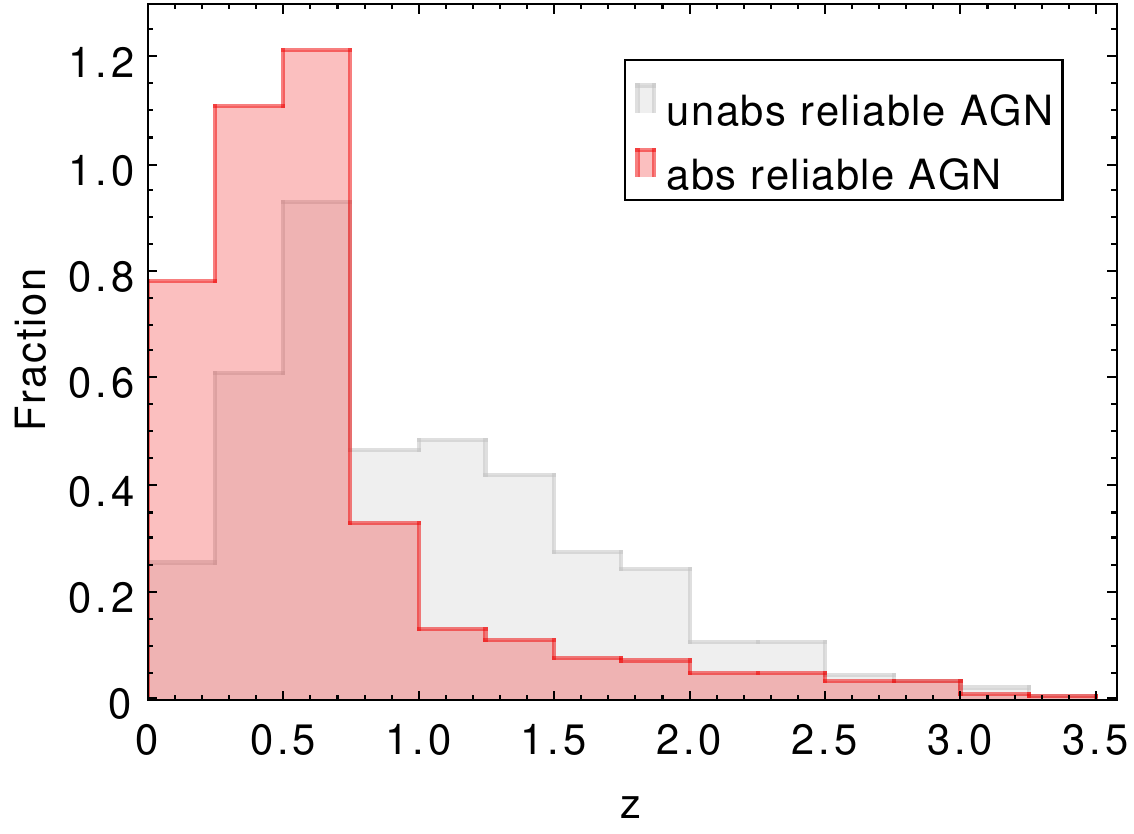}
  \includegraphics[width=0.49\linewidth, height=4cm]{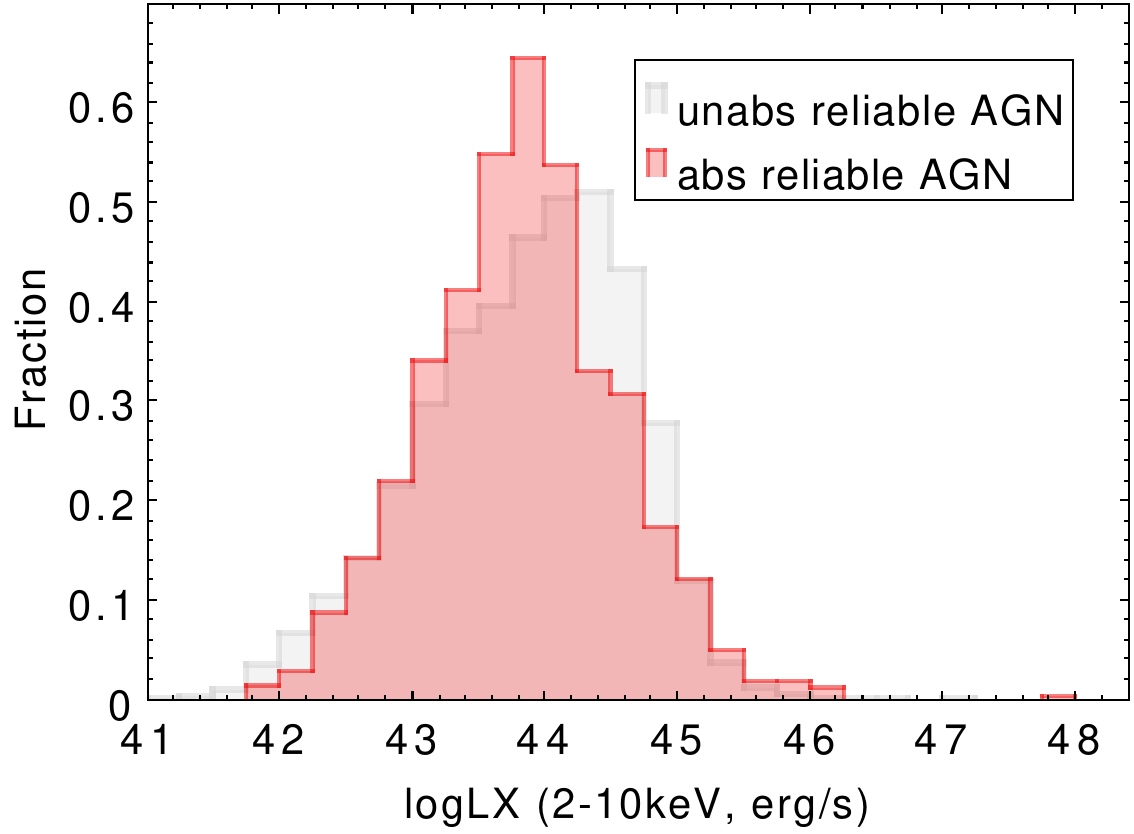}
  \caption{%
    Top panel: decimal logarithm of the $2-10\,{\rm keV}$ rest-frame intrinsic
    luminosity versus the redshift of the reliable \ac{AGN} sample. Grey dots
    corresponds to the \myunabsagn{} and red dots correspond to the
    \myabsagn{}\@. Note that due to the discreteness of the zmode0 from tpz (see
    the text for the details), the redshift value plotted here is offset by a
    random number between $-0.05$ and $0.05$. Bottom panel: histogram of the
    redshift (left) and luminosity (right) of the \textit{Good fit} and reliable
    \myabsagn{} and \myunabsagn{} samples.
  }
  \label{fig:logLXvsz}
\end{figure}

\begin{figure}[htbp]
  \centering
  \includegraphics[width=0.85\linewidth, height=6.5cm]{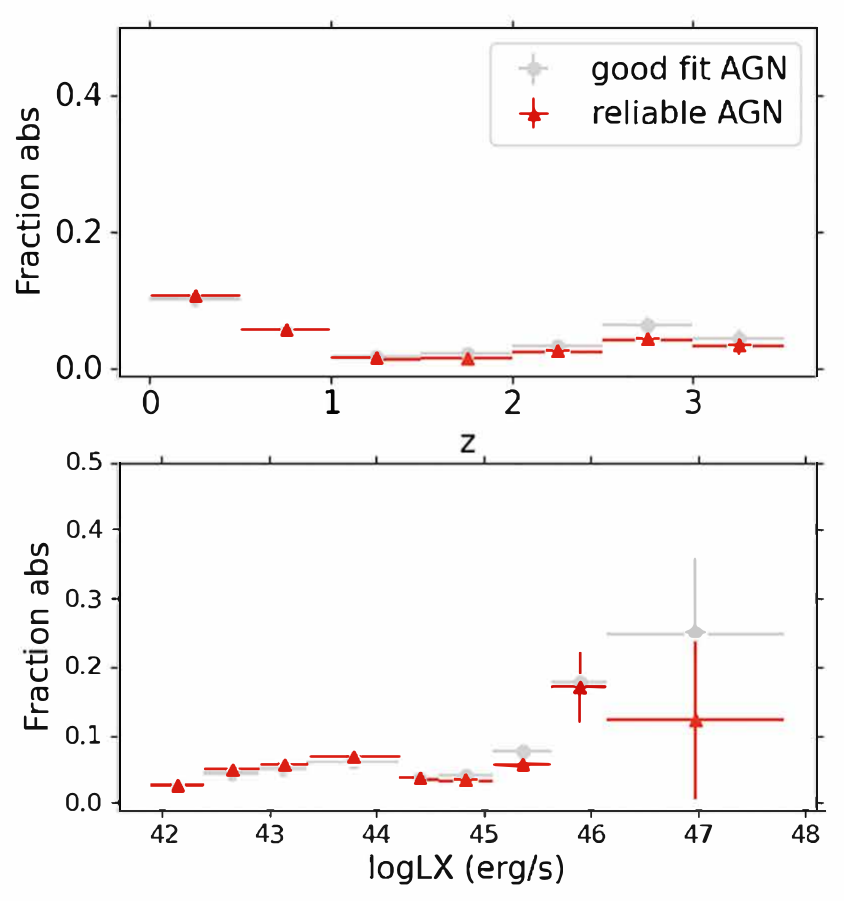}
  \caption{%
    Fraction of absorbed \acp{AGN} as a function of redshift (top) and luminosity
    (bottom) both for the \textit{Good fit} \ac{AGN} sample (grey dots) and for the
    reliable \ac{AGN} sample (red triangles).%
  }
  \label{fig:fabszlogLX}
\end{figure}

In Fig.~\ref{fig:logLXvsz} we show the distributions of the redshifts and
luminosities of the reliable \myabsagn{} and \myunabsagn{} samples. The
redshift distributions are significantly different (the KS $p$-value is
essentially 0): the medians are 0.53 and 0.83, respectively, with \myabsagn{}
concentrating at $z<1$ ($>85$\% are below that value), while \myunabsagn{} have
a higher fraction of sources at higher z ($>57$\% are above that value). The
luminosity distributions are also significantly different (KS $p$-value
corresponding to $>6$-sigma) but the differences are quantitatively small, with
medians of $43.82$ and $43.95$, respectively. The higher incidence of absorbed
sources at lower redshifts may be a selection effect due to, e.g., the lower
number of counts detected from absorbed sources, since then their spectra may
not be extracted by the \XMM{} pipeline. Additionally, the effect of absorption
is higher at lower energies, and this range moves out of the \XMM{} range as
the redshift increases, making the effect of absorption more difficult to
detect at higher z \citep[as discussed e.g.\ by][]{Marchesi2016}. This feature
is also seen in the top panel of Fig.~\ref{fig:fabszlogLX}, where the first two
redshift bins are higher than the rest. Similar properties can be found using
the \mylognhxxiii{} \myabsagn{} definition and the more restrictive
\myabsunabsundet{} \myunabsagn{} definition.

We show in Fig.~\ref{fig:fabszlogLX} the fraction of absorbed sources defined
as $N_{\rm abs}/(N_{\rm abs} + N_{\rm unabs})$ as a function of redshift (top
panel) and log luminosity (bottom panel). We have estimated the error bars
using the expression for the binomial distribution with high number of sources.
We have used Bayesian blocks \citep{Scargle2013} to define the log
luminosity bins.\footnote{%
This has not been possible for the redshift, since the photometric redshifts
used in this work are zmode0 from tpz with step size of 0.1, and
this confused the algorithm. We have used instead a fixed width of 0.5 for the
bins.%
}
As in R21 (their Fig.~6), we find no evidence of a dependence of the fraction
of absorbed sources with redshift (except for the first two bins in z, probably
due to selections effects, see above), also getting similar fractions and
little difference between the \textit{Good fit} and reliable samples. The picture
is also similar when looking for a dependence on luminosity, no significant one
is found: the highest luminosity bin for the \textit{Good fit} \ac{AGN} sample is
higher than for the reliable sample, but compatible with it, within errors. The
latter is also compatible with the lower luminosity bins.
Estimating fractions with the higher threshold for the definition of
\myabsagn{}, \mylognhxxiii{} produces a lower fraction of absorbed sources (as
expected, since the higher threshold is more stringent). The qualitative
behavior of the fractions as a function of redshift and luminosity are the
same. The introduction of the \myundetagn{} category does not affect the
default results, since the fraction of absorbed \acp{AGN} does not change.

In contrast with this constant and low fractions, other studies with similar
luminosity medians and similar or higher redshifts, but correcting for the
selection effects discussed at the beginning of this section and using more
sophisticated spectral models
\citep[e.g.][]{Peca2023,Vijarnwannaluk2022,Buchner2015,Signorini2023,Aird2015,Pouliasis2024},
find much higher absorbed fractions $\sim 0.6$, generally increasing with
redshift and decreasing with luminosity, except for \citet{Pouliasis2024}, who
find a constant fraction with redshift between local \citep{Boorman2025}
samples and their redshift $3-6$ sample.

\begin{figure}[htbp]
  \centering
  \includegraphics[width=0.65\linewidth]{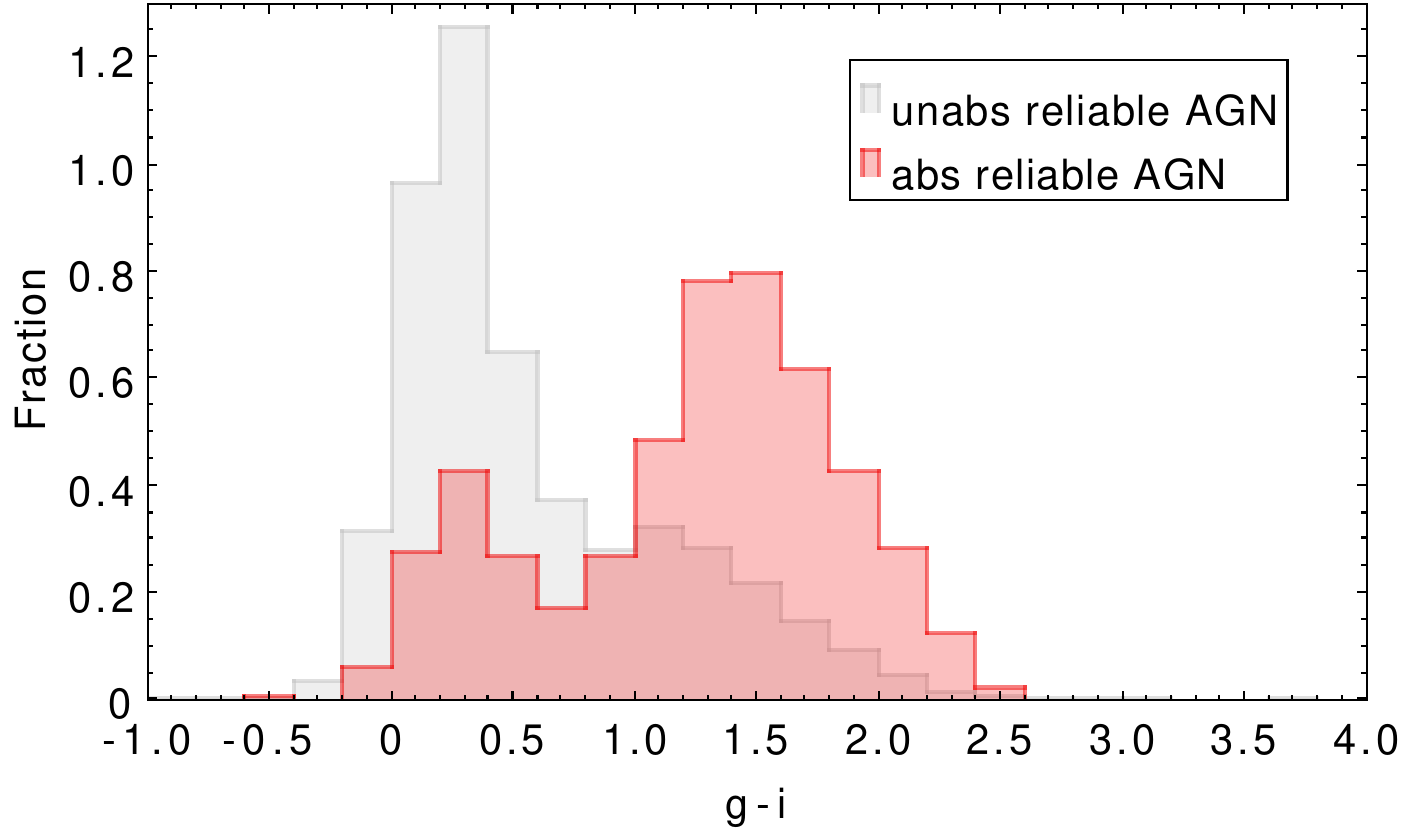}
  \includegraphics[width=0.65\linewidth]{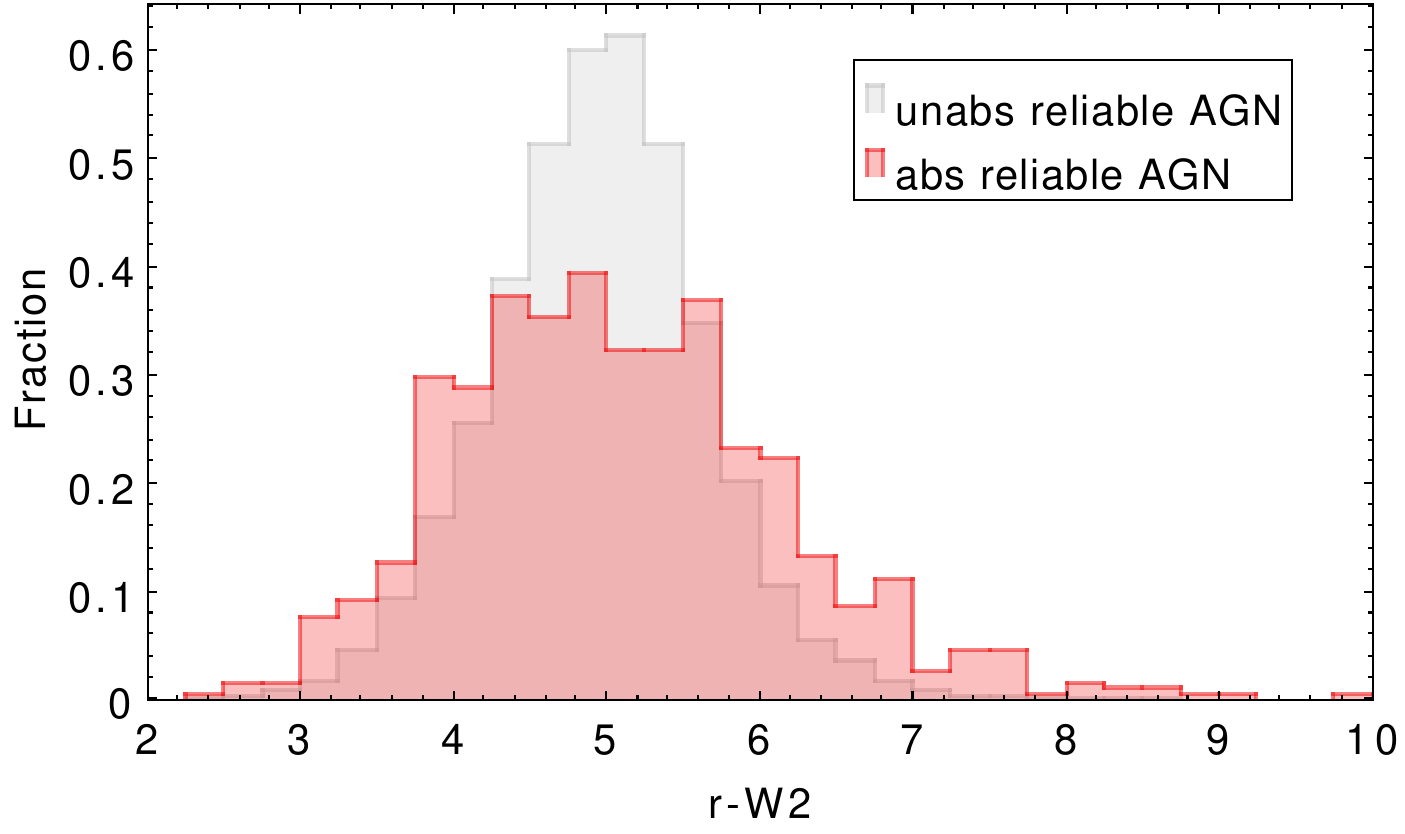}
  \includegraphics[width=0.65\linewidth]{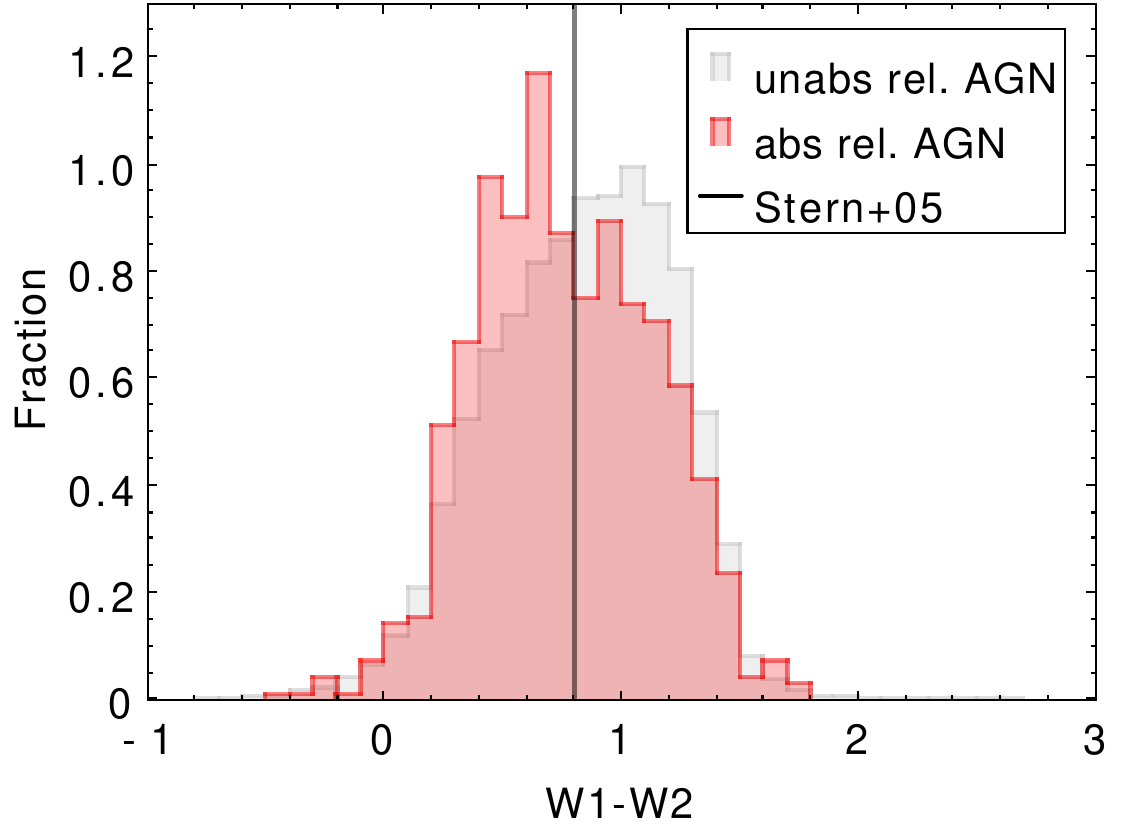}
  \caption{%
    Distribution of $g-i$ (top), $r-W2$ (middle) and $W1-W2$ (bottom) colours of
    our reliable \ac{AGN} samples, with \myabsagn{} in red and \myunabsagn{} in
    grey. In the bottom plot we also show the simple $W1-W2$ colour criterion to
    select \acp{AGN} from \cite{Stern2005}.%
  }
  \label{fig:colorhist}
\end{figure}

\begin{figure}[htbp]
  \centering
  \includegraphics[width=0.9\linewidth, height=5cm]{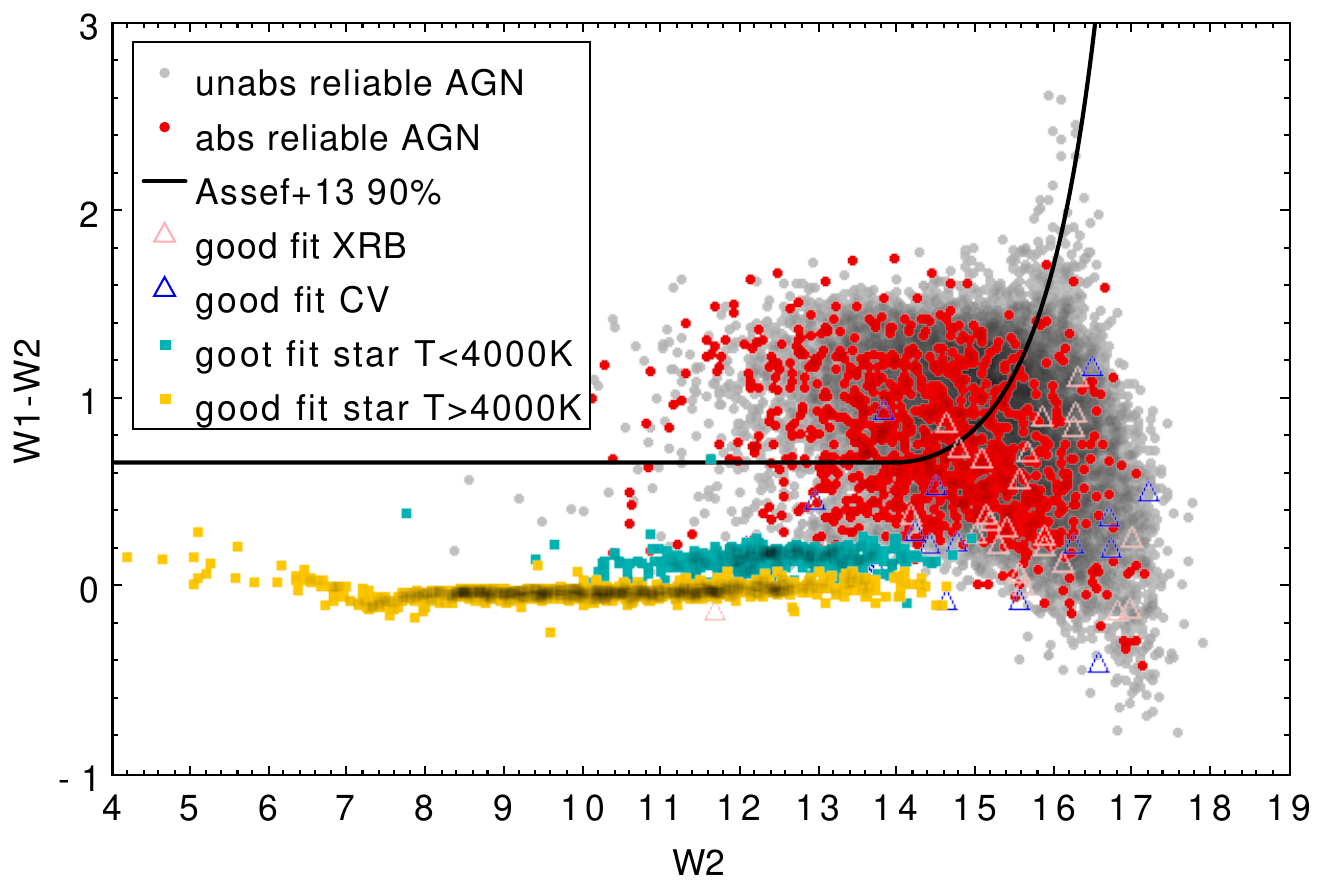}
  \caption{%
    WISE $W1-W2$ colour vs $W1$ for reliable \myabsagn{} (red dots),
    \myunabsagn{} (grey dots), \acp{XRB} (pink triangles), \acp{CV}
    (blue triangles), \mylowt{} stars (cyan squares), and \myhight{} stars
    (gold squares). The locus of \acp{AGN} (above the black line) of
    \citet{Assef2013} is also shown.%
  }
  \label{fig:W1W2}
\end{figure}

Mid-infrared (MIR) colours are often used for identifying \acp{AGN}. In
particular, the Wide-field Infrared Survey Explorer (WISE) $W1$ (3.4~$\mu$m),
$W2$ (4.6~$\mu$m), $W3$ (12~$\mu$m) and $W4$ (22~$\mu$m) bands
\citep[e.g.][]{Stern2012,Mateos2012,Assef2013,Glikman2018}. Galaxies are
generally expected to exhibit bluer colours than \acp{AGN} in the MIR regime.
The bottom panel of Fig.~\ref{fig:colorhist} shows the $W1-W2$ colours of our
default sample. There is considerable overlap between the $W1-W2$ colours of
\myabsagn{} and \myunabsagn{}, and the differences in the medians are small
compared to the dispersion (0.75/0.86 for \myabsagn{}/\myunabsagn{}),
consistent with the expectation that the MIR emission originates from the torus
and is independent of the inclination \citep[see][and references
therein]{Padovani2017}. The initial simple $W1-W2$ criterion of
\cite{Stern2012} has been updated by \cite{Assef2013} introducing a dependence
on the $W2$ magnitude. In Fig.~\ref{fig:W1W2}, we present the $W1-W2$ colour
distribution vs $W2$ for our default reliable \myabsagn{}/\myunabsagn{}
definition, \acp{XRB}, \acp{CV} and stars with GAIA detections. We also show in
that figure one such criterion, showing that it is unable to select about half
the \acp{AGN} in our sample, both for the \myabsagn{} and \myunabsagn{}, as
found by R21. A similar conclusion is reached if the criterion of
\cite{Assef2018} is used instead. Using the \mylognhxxiii{} and
\myabsunabsundet{} limits also provide similar results. Our results and those
of R21 are in agreement with \cite{Hickox2017}, who show that luminous quasars
can be effectively selected using simple MIR colour criteria, but those
criteria fail for heavily obscured and lower luminosity \acp{AGN}.

Other optical and mixed optical-MIR colours are commonly used to select \acp{AGN}.
The middle panel of Fig.~\ref{fig:colorhist} shows $r-W2$ for our reliable \ac{AGN}
sample: they are centered in similar values, but the \myabsagn{} sample is
wider, with KS $p$-values $\sim 0$. \cite{Yan2013} propose that $r-W2>6$ allows
for selecting obscured \acp{AGN}, combined with pure MIR diagnostics, arguing
that the $r$ band will suffer from extinction more strongly than $W2$. R21 find
a slight tendency for \myabsagn{} to have a stronger $r-W2>6$ tail than
\myunabsagn{}, which is even weaker in our sample. Those authors offer an
explanation: \cite{Hickox2017} show that the $r-W2$ criterion is only actually
effective for $z>1$, while most of our (and R21) \myabsagn{} are below that
redshift. The more strict definition of \myabsagn{} using \mylognhxxiii{}
reduces the number of $z>1$ \myabsagn{}, and thus the $r-W2>6$ tail for
\myabsagn{}, even more. The more stringent definition of \myabsunabsundet{}
produces equivalent results to those of our standard definition.

\begin{figure}[htbp]
  \centering
  \includegraphics[width=0.75\linewidth]{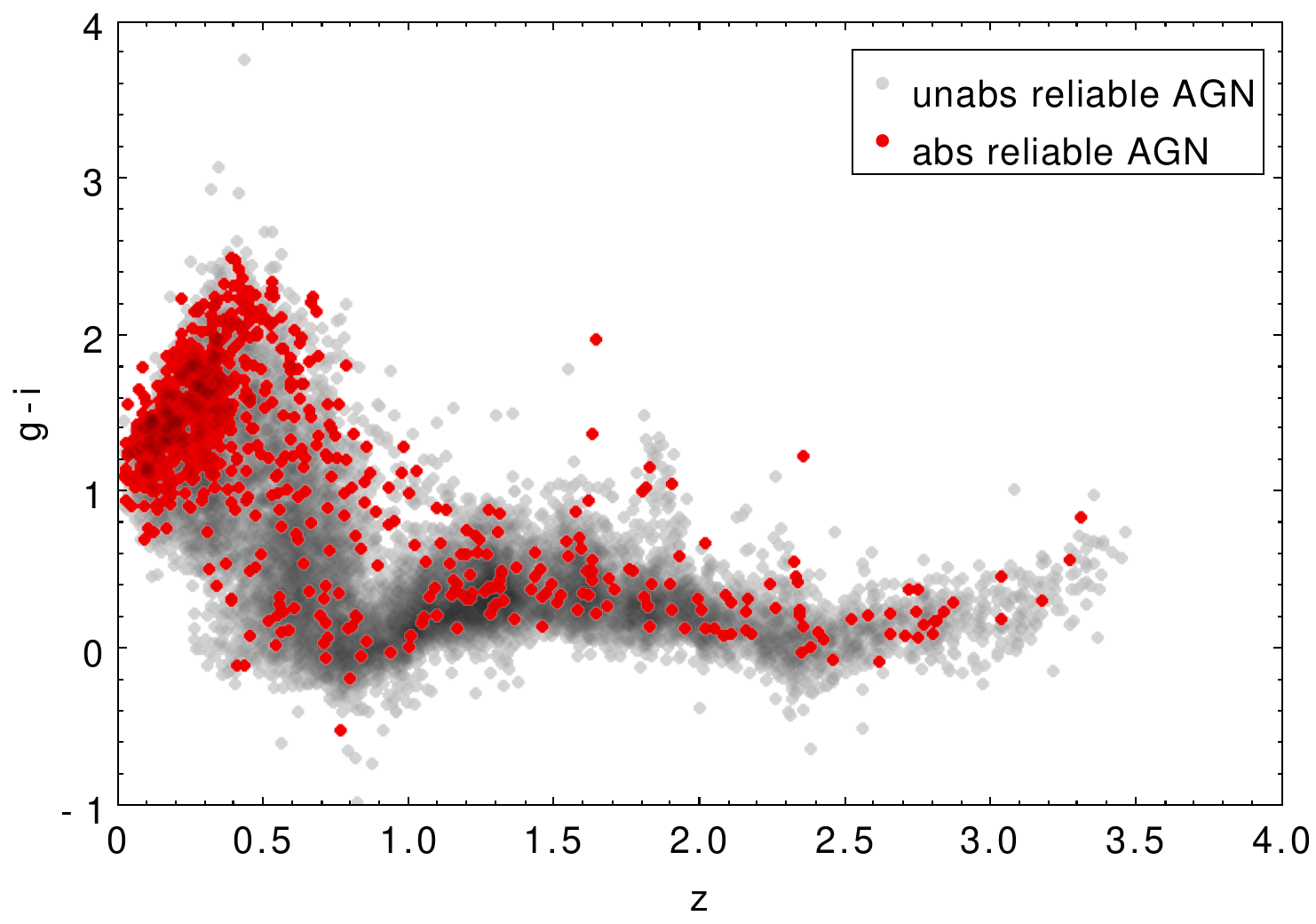}
  \caption{%
		The colour $g-i$ versus redshift for our reliable \ac{AGN} sample:
		\myabsagn{} in red and \myunabsagn{} in grey.%
  }
  \label{fig:giz}
\end{figure}

We also display in Fig.~\ref{fig:colorhist} (top) the distribution of $g-i$ for
our reliable \ac{AGN} sample. Since the presence of X-ray absorption is often
accompanied by optical obscuration, and the effect of the latter is more
pronounced at optical/UV wavelengths, it is not surprising at first view that
\myabsagn{} peak at redder $g-i$ colours than \myunabsagn{}, but the presence of
a second peak in the \myabsagn{} distribution at $g-i \sim 0.3$ and the
significant tail with $g-i>1$ for \myunabsagn{} reveal a more complex story. As
can be appreciated in Fig.~\ref{fig:giz}, redshift has a strong effect on
$g-i$: most of the $g-i>1$ sources are at $z<1$, and most of the $g-i<1$
sources are at $z>1$, with no clear difference between \myabsagn{} and
\myunabsagn{} in the latter redshift range. If we restrict the analysis to
$z<1$ there is a clear preponderance of \myabsagn{} at $g-i>1$, with
\myunabsagn{} more spread in the $g-i$ range of ${\sim} 0.01$ to ${\sim} 1.7$.
Quantitatively, the median of $g-i$ for $z<1$ \myunabsagn{} is 0.84, while
$>90$\% of \myabsagn{} in the same redshift range have $g-i>0.84$. The colour
$g-i$ allows selecting \myabsagn{} only at $z<1$ and with a strong mixture of
\myunabsagn{}. We reach a similar conclusion with the \mylognhxxiii{} limits.

Comparable results are again obtained with the more tight limits for
\myabsunabsundet{}: now $>93$\% of $z<1$ \myabsagn{} are above the median value
for genuine \myunabsagn{} $g-i=0.71$. It is interesting to note that this
median is lower than for the default and \mylognhxxiii{} limits, and it is
plainly lower than for \myundetagn{}, which have a median $g-i=1.25$, $>74$\%
of them having $g-i>0.71$. The more stringent \myabsunabsundet{} limits
patently make a difference in the $g-i$ colour between \myundetagn{} and genuine
\myunabsagn{} at $z<1$

To summarise our conclusions about the use of optical, MIR or mixed optical-MIR
colours to select absorbed \acp{AGN}: $W1-W2$ and $r-W2$ are not successful in
separating absorbed \acp{AGN} in our samples, not even when restricting the selection
to $z>1$ or using higher values of the fitted column density $N_{\rm H}$ to
define absorbed sources. The colour $g-i$ is a bit more successful at $z<1$,
with absorbed \acp{AGN} having mostly $g-i > 0.8-1$, but with a strong contamination
of unabsorbed \acp{AGN} (about half the unabsorbed \ac{AGN} colours are above
that limit). Similar conclusions are reached if a higher limit is used to
define absorbed sources or if we take into account the uncertainties on the
fitted values of $N_{\rm H}$ to define unabsorbed sources.

In contrast, \acp{CV} and \acp{XRB} tend to agree with the \cite{Assef2013} criterion in
Fig.~\ref{fig:W1W2}, their MIR colours being closer to stellar than \ac{AGN},
unsurprisingly. Stars tend to occupy a pair of roughly horizontal and parallel
branches in this diagram, well below the \ac{AGN} boundary, the top branch
corresponding to \mylowt{} stars and the bottom branch corresponding to
\myhight{} stars. The $W1-W2$ colours of stars have already been discussed by
\cite{Nikutta2014}, who showed that main sequence stars without significant
circumstellar IR excess (`naked' stars) have a range of $W1-W2$ colours, with
Vega-like hot stars having $W1-W2 \sim 0$ and cooler stars extending up to
$0.2-0.3$ (see their Fig.~7). The two branches disappear when the absolute $W2$
magnitude instead of the apparent one is plotted in the horizontal axis (using
the GAIA distance), with the top branch becoming a continuation of the lower
branch towards higher absolute magnitudes (lower luminosities) and higher
$W1-W2$ colour. As stated above, the \XMM{} spectra of the objects classified as
stars have been fitted with a single APEC model in C4. Given the above
dependence on GAIA $T_{\rm eff}$, we have checked whether there is any
correlation between any of those parameters (temperature $kT$, flux, line of
sight column density $N_{\rm H}$ and $T_{\rm eff}$), finding none. This is not
surprising, since X-ray emission in main sequence stars is mainly from the
corona of the star, which is thought to be heated by magnetic reconnection,
which is powered by a dynamo effect in the outer layers of the star, but well
below the surface where the GAIA $T_{\rm eff}$ is measured.

\section{Conclusions}
\label{sec:conclusions}

In this study, we employed X-ray detections, sources and spectra derived from
the 4XMM-DR11 and 4XMM-DR11s catalogues, subjecting them to fitting procedures
employing both simple and physically-motivated models. Our analysis yielded the
creation of four distinct catalogues. The first catalogue (C1) utilised the
4XMM-DR11 detections, employing a simple absorbed power-law model to analyze the
extracted X-ray spectra in the \XMM{} pipeline. This provided insights into the
distribution of flux, \ac{IIN}, hydrogen column density ($N_{\rm H}$), and
photon index ($\Gamma$). The second catalogue (C2) presented results from
applying both absorbed power-law and absorbed blackbody models to merged spectra
from the stacked sources from the 4XMM-DR11s catalogue, providing additional
information such as blackbody temperature. The third catalogue (C3) uses the
count rates from the source detection as low resolution `spectra', fitting
simple absorbed power-laws to them, expanding the results of C1 to all the
detections, not only those with sufficient net counts to have extracted spectra
in the \XMM{} pipeline. The fourth catalogue (C4) concentrated on \acp{AGN}, \acp{XRB},
\acp{CV}, and stars, incorporating classifications from \cite{Tranin2022} and
fitting a combination of the detection spectra in C1 and the merged spectra in
C2. For each class of sources a distinct physically-motivated model was fitted,
allowing an easier physical interpretation of the results. The catalogue
entries include median and mode values for all calculated parameters, along
with the 5th and 95th percentiles and the narrowest 90 per cent interval.
Throughout this work, we report results using the mode values of the
parameters. The main conclusions of our analysis are summarised below:
\begin{itemize}

  \item[$\bullet$] Sources in C1 present a median flux of $\mylogfx = -13.44$,
    with \ac{IIN} mostly around $1.05$. Spectral parameters include median $\mylognh
    = 21.26$ and $\Gamma=1.95$ (probably because the catalogue source
    population is dominated by \acp{AGN}, which have typically similar values of
    their photon index). Some sources show a secondary tail in $\Gamma$ at high
    values, likely from attempting to fit thermal emission with an absorbed
    power law.

  \item[$\bullet$] In C2, absorbed power-law and absorbed blackbody models
    provide flux measurements with a median difference of ${\sim} 0.2$~dex,
    perhaps because the former is a better model for the majority of the
    sources (\acp{AGN}). There is a good agreement between the \ac{IIN}
    distributions of the two models. Blackbody temperatures generally below
    $3\,{\rm keV}$ are found for the majority of sources.

  \item[$\bullet$] The fit to all detections in C3 allows extending the simple
    absorbed power-law modelling to all detections in 4XMM-DR11, providing some
    spectral information from fainter sources than available in C1. The derived
    fluxes are compatible between the two approaches.

  \item[$\bullet$] In C4, the results show a strong agreement in flux
    measurements with C1 for \acp{AGN}, stars, \acp{XRB} and \acp{CV}. The \ac{IIN} and
    $\Gamma$ comparisons between C1 and C4 exhibit good consistency. The
    comparison of the flux and spectral parameters between sources classified
    as \acp{AGN} in the C4 with sources in C2 fitted with an absorbed power-law shows
    good agreement. The comparison between the C4 sources classified as \acp{XRB}
    and the blackbody model measurements in the C2 catalogue, shows also a good
    agreement regarding the \ac{IIN} parameter, although the $f_{\rm X}$
    calculations have a mean difference of ${\sim} 0.2$ dex. Primarily, this
    difference is attributed to the enhanced fits of sources in the C4 dataset,
    as evidenced by the respective $p$-values. The finding also aligns with the
    expectation that a power-law component, in conjunction with a blackbody,
    provides a better fit for \acp{XRB} than a blackbody alone.

  \item[$\bullet$] After following the analysis of \cite{Ruiz2021}, we share
    most of their conclusions, with the important caveat that the
    \textit{reliable} sample of \acp{AGN} as presented in Sect.~\ref{sec:science} is
    not flux-limited or complete in any sense: there is no significant change
    of the fraction of X-ray absorbed \acp{AGN} (those with the lower end of the 90\%
    uncertainty of the column density $>10^{22}\,{\rm cm}^{-2}$) with either
    redshift or intrinsic luminosity. The $W1-W2$ or $r-W2$ optical and MIR
    colours are of limited value to separate X-ray absorbed and unabsorbed \acp{AGN},
    even when restricting the samples to $z>1$. The optical $g-i$ colour is a
    bit more successful, especially when restricting the analysis to $z<1$.
    This does not change if a higher limit is used for the definition of
    absorbed \acp{AGN} (lower end $>10^{23}\,{\rm cm}^{-2}$) or if a more stringent
    definition of unabsorbed \acp{AGN} is used (upper end of the 90\% uncertainty on
    the column density $<10^{22}\,{\rm cm}^{-2})$.

  \item[$\bullet$] The $W1-W2$ vs $W2$ distribution of stars, \acp{XRB}, and
    \acp{CV}, are within the expected boundaries for non-\acp{AGN}.
    Stars show two branches corresponding to higher temperature ($T_{\rm
    eff}>4\,000\,{\rm K}$) main sequence stars without significant
    circumstellar IR excess, and to lower luminosity cooler stars ($T_{\rm
    eff}<4\,000\,{\rm K}$), respectively.

\end{itemize}
The four catalogues contribute valuable insights into the X-ray properties of
celestial sources, demonstrating their utility in understanding the diverse
phenomena observed by \XMM{} over the past two decades. The methodologies and
comparisons presented enhance the reliability of the catalogues and pave the
way for further scientific applications.

\begin{acknowledgements}

This project has received funding from the European Union's Horizon 2020
research and innovation program under grant agreement no. 101004168, the
\xmmiiathena{} project. GM acknowledges funding from grant PID2021-122955OB-C41
funded by MCIN/AEI/10.13039/501100011033 and by ERDF A way of making Europe. JB
acknowledges support from Centre National d’Études Spatiales (CNES) for their
outstanding support for the SSC activities. NW is grateful for support from the
CNES for this work. HS acknowledges support from the ``Big Bang to Big Data''
(B3D) project, the NRW cluster for data-intensive radio astronomy, funded by
the state of North Rhine-Westphalia as part of the \textit{Profiling 2020}
programme. This research has made use of data obtained from the 4XMM \XMM{}
serendipitous source catalogue compiled by the 10 institutes of the \XMM{}
Survey Science Centre selected by ESA\@. This work has made use of data from
the European Space Agency (ESA) mission \textit{Gaia}
(\url{https://www.cosmos.esa.int/gaia}), processed by the \textit{Gaia} Data
Processing and Analysis Consortium (DPAC,
\url{https://www.cosmos.esa.int/web/gaia/dpac/consortium}). Funding for the
DPAC has been provided by national institutions, in particular the institutions
participating in the \textit{Gaia} Multilateral Agreement.

\end{acknowledgements}

\bibliographystyle{aa.bst}
\bibliography{mybib.bib}

\appendix

\section{Comparison of parameter estimates between catalogues}
\label{appendix_cross_checks}

In this section, we compare the measurements of various parameters across the
four catalogues, as well as between different models used within each
catalogue. Our goal is to evaluate their consistency and identify any
systematic differences.

\subsection{Comparison of flux and IIN Estimates in C2: power-law vs.\ blackbody models}
\label{appendix_c2}

\begin{figure*}[htpb]
  \centering
  \includegraphics[width=0.8\linewidth, height=5cm]{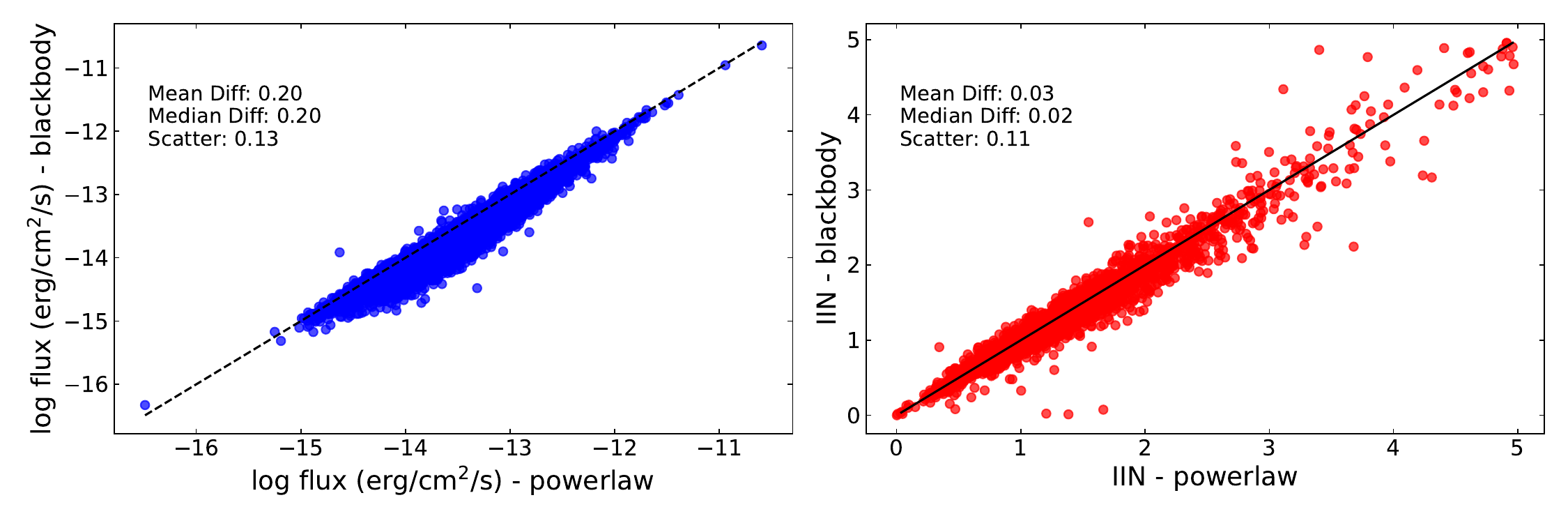}
  \caption{%
    Comparison between the calculations of the power-law and the blackbody
    models, for the $f_{\rm X}$ (left panel) and IIN (right panel), for the
    sources included in the \textit{Good fit} sample of C2.%
  }
  \label{fig:c2_comp_appendix}
\end{figure*}

Fig.~\ref{fig:c2_comp_appendix} presents a comparison between the $f_{\rm X}$
(left panel) and \ac{IIN} (right panel) calculations of the power-law and the
blackbody models. There is a good correlation of the fluxes obtained by the two
models. The power-law model tends to calculate higher fluxes compared to the
blackbody model. The difference of the two flux calculations has a mean value
of 0.21 (median value of 0.20) and a scatter of 0.13. The reason for this
difference is perhaps that most X-ray sources are expected to be \acp{AGN}, for which
an absorbed power-law is in principle a better fit than a blackbody. A very good
agreement is found regarding the \ac{IIN} calculations of the two models, with a
mean difference of 0.03 and a scatter of 0.11.

\subsection{C1–C3 Catalogue Comparison}
\label{appendix_c1_c3}

\begin{figure*}[htbp]
  \centering
  \includegraphics[width=0.9\linewidth, height=11cm]{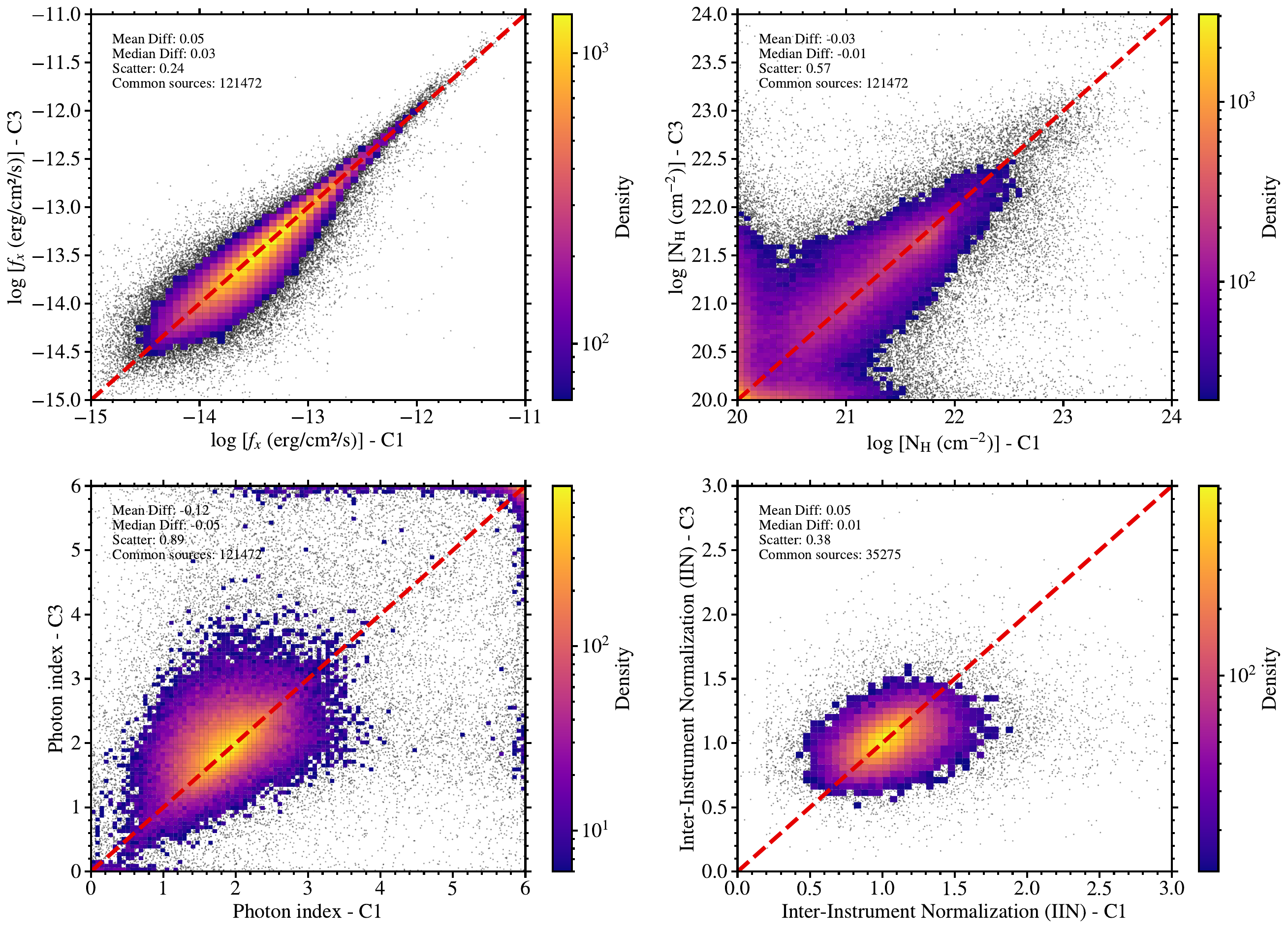}
  \caption{%
    Comparison of $f_{\rm X}$ (top-left panel), $N_{\rm H}$ (top-right
    panel), photon index (bottom-left panel) and \ac{IIN} (bottom-right panel)
    calculations between the C3 catalogue and C1, using the \textit{Good fit}
    subsets of the two catalogues. Each panel displays the mean, median
    difference, and scatter.%
  }
  \label{fig:C3_C1_comp_appendix}
\end{figure*}

Figure~\ref{fig:C3_C1_comp_appendix} shows a comparison of the calculated
values for $f_{\rm X}$, $N_{\rm H}$, $\Gamma$, and \ac{IIN} between the C3 and C1
catalogues. This comparison helps us evaluate the differences between using
count rate spectra (C3) and applying proper spectral fitting (C1).

Our findings indicate that $f_{\rm X}$ results from the two methods are in good
agreement, with the mean and median differences being 0.05 and 0.02,
respectively, and a scatter of 0.24. The correlation for $N_{\rm H}$ and
$\Gamma$ is also decent, although the scatter is notably larger. The
distribution plots for these parameters show a high density of sources at the
edges of the parameter space. This pattern emerges because, in some cases, one
method provides a well-defined posterior distribution for a parameter, while
the other method does not. Sources with poorly constrained posteriors can be
identified in the catalogue by their large credible intervals for the parameter
in question. Using a basic model, we attribute poorly constrained posteriors to
low count statistics, where the number of detected counts is similar to the
background noise level. Regarding the \ac{IIN} parameter, most measurements cluster
around a value of one, as expected, but the correlation is weaker compared to
other parameters. It is worth noting that in the C1 catalogue, the \mos{}
spectra were merged into a single spectrum, which could influence the \ac{IIN}
values.

In summary, the use of count rate spectra for estimating spectral parameters
can be a viable approach for population studies, as long as the catalogue is
properly filtered. However, for examining individual sources, extracting and
fitting a detailed spectrum remains the preferred method.

\subsection{Comparison of C4 with Other Catalogues}
\label{appendix_c4_vs_all}

\begin{figure*}[htbp]
  \centering
  \includegraphics[width=0.8\linewidth, height=11cm]{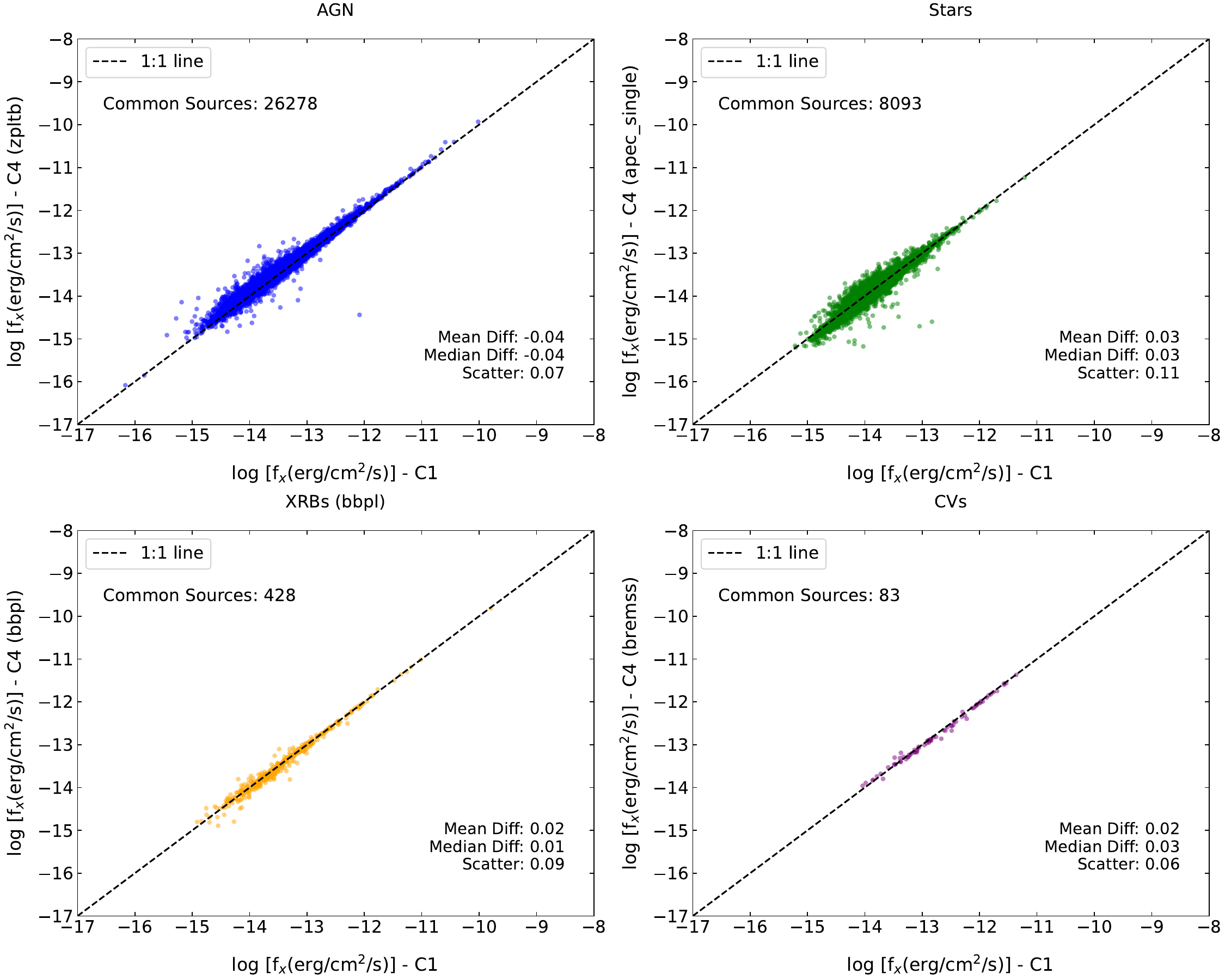}
  \caption{%
    Comparison of $f_{\rm X}$ calculations between the C4 catalogue and C1,
    categorised by source type in the C4 catalogue, for sources included in the
    \textit{Good fit} samples. Each panel displays the mean, median difference,
    and scatter.
  }
  \label{fig:fx_C4_vs_c1_appendix}
\end{figure*}

\begin{figure*}[htbp]
  \centering
  \includegraphics[width=0.8\linewidth]{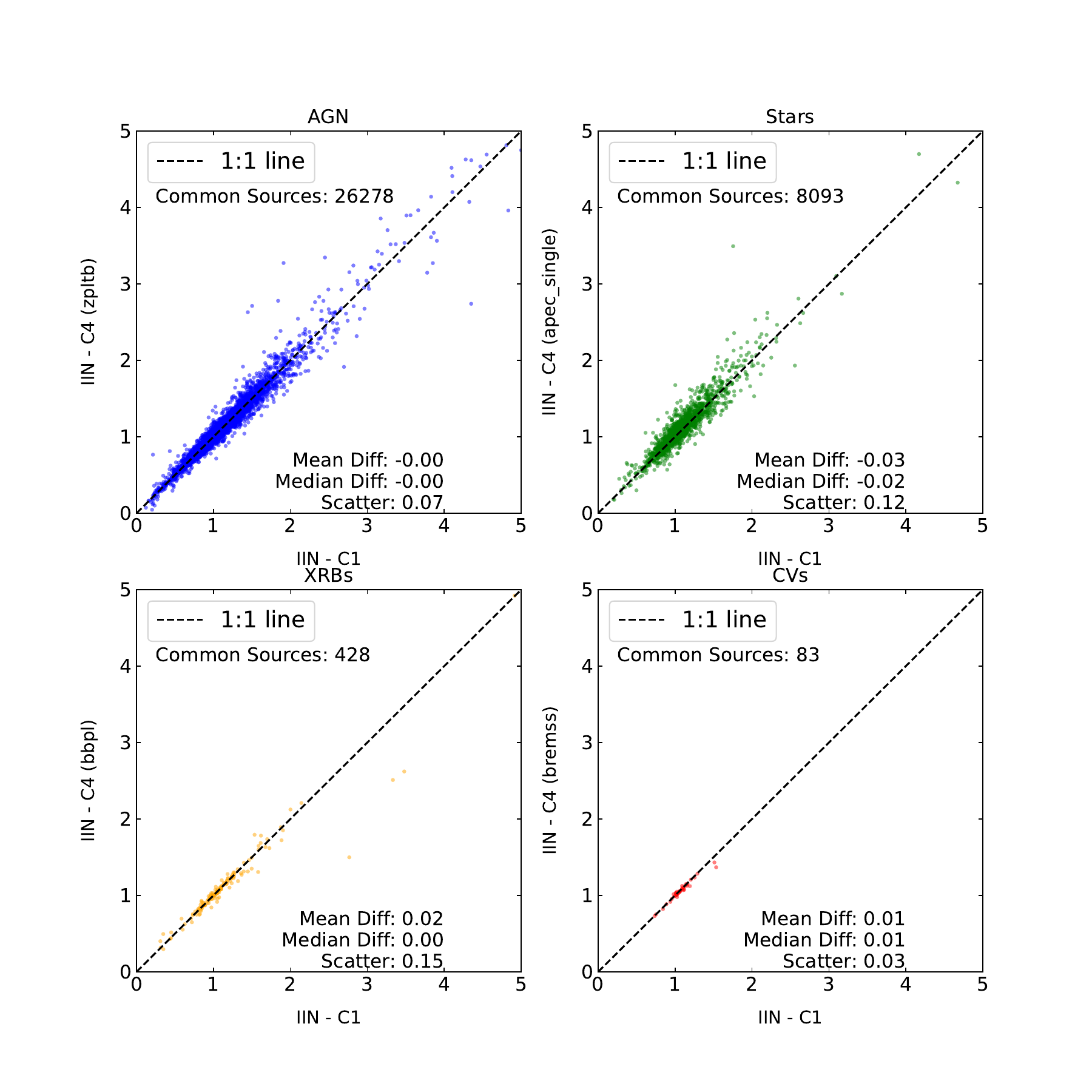}
  \caption{%
    Comparison of \ac{IIN} calculations between the C4 catalogue and C1, categorised
    by source type in the C4 catalogue, for sources included in the
    \textit{Good fit} samples. Each panel displays the mean, median difference,
    and scatter.
  }
  \label{fig:iin_C4_vs_c1_appendix}
\end{figure*}

\begin{figure*}[htbp]
  \centering
  \includegraphics[width=0.95\linewidth, height=7.5cm]{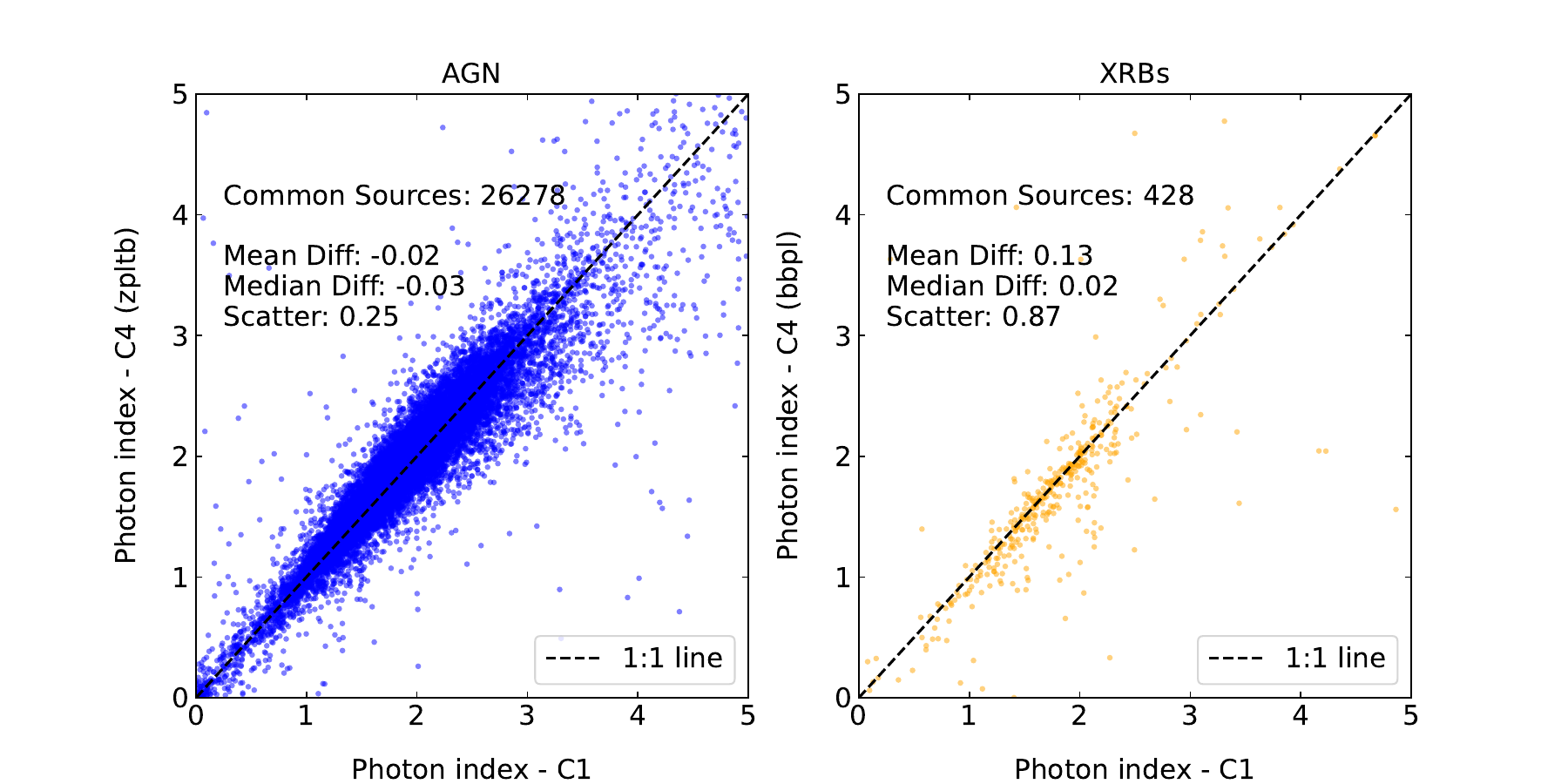}
  \caption{%
    Comparison of $\Gamma$ calculations between the C4 catalogue and C1,
    categorised by source type in the C4 catalogue, for sources included in the
    \textit{Good fit} samples. Each panel displays the mean, median difference,
    and scatter.
  }
  \label{fig:gamma_C4_vs_c1_appendix}
\end{figure*}

\begin{figure*}[htbp]
  \centering
  \includegraphics[width=\linewidth, height=7.5cm]{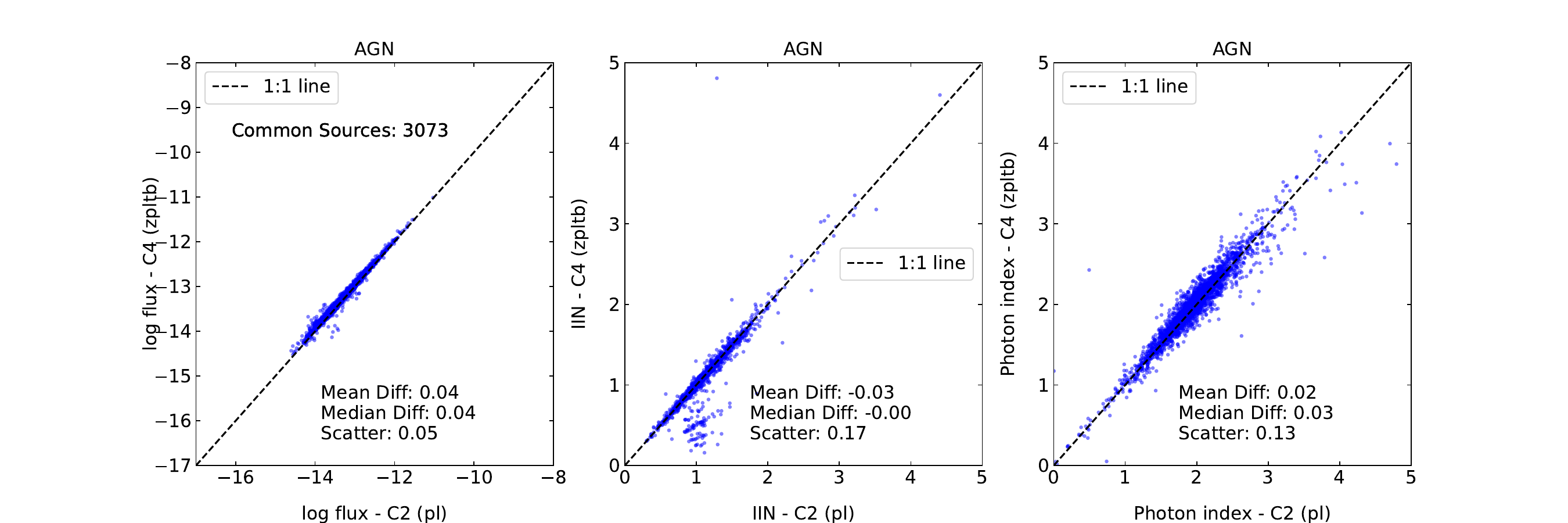}
  \caption{%
    Comparison of the $f_{\rm X}$ (left panel), \ac{IIN} (middle panel), and
    $\Gamma$ (right panel) measurements for sources within the C4 catalogue,
    where the \zpltb{} model is utilised for spectral fitting (i.e.\ \acp{AGN}),
    with corresponding measurements from C2, employing a power-law model for
    spectral fitting. Each panel displays the mean, median difference, and
    scatter. Sources included in the \textit{Good fit} samples are presented.
  }
  \label{fig:C4_vs_c2_pl_appendix}
\end{figure*}

\begin{figure*}[htbp]
  \centering
  \includegraphics[width=1.00\linewidth, height=8cm]{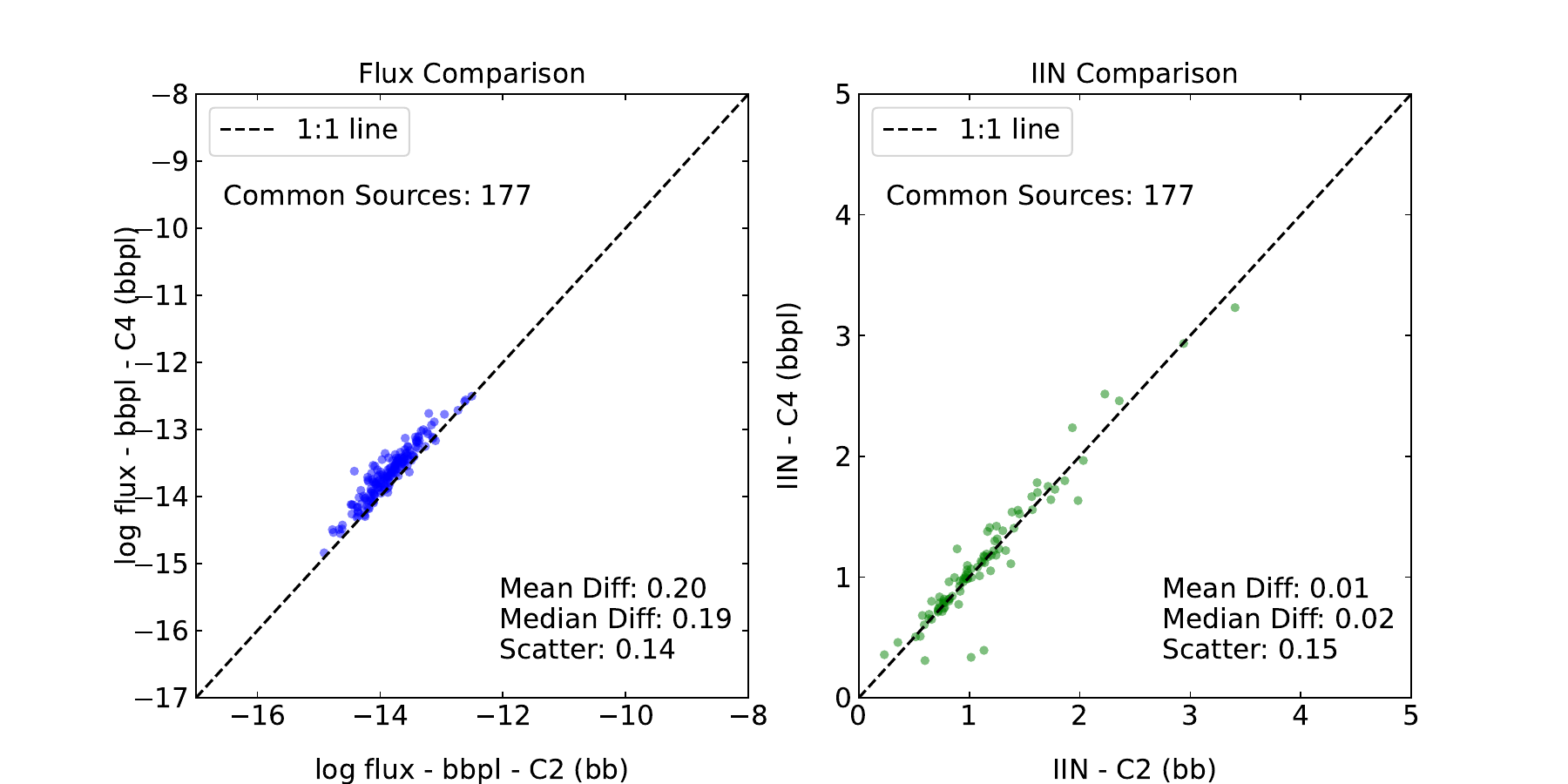}
  \caption{%
    Comparison of $f_{\rm X}$ and \ac{IIN} calculations for sources within the C4
    catalogue, where their X-ray spectra are fitted with a \bbpl{} model (i.e.,
    \acp{XRB}), and sources in the C2 catalogue fitted with a blackbody model for
    their spectral analysis. Each panel displays the mean, median difference,
    and scatter. Sources included in the \textit{Good fit} samples are
    presented.
  }
  \label{fig:C4_vs_c2_bb_appendix}
\end{figure*}

Figure~\ref{fig:fx_C4_vs_c1_appendix} presents the comparison of $f_{\rm X}$
calculations between C4 and C1, categorised by source type in the C4 catalogue.
Each panel displays the mean, median difference, and scatter. There is strong
agreement between the $f_{\rm X}$ calculations in both catalogues. Likewise,
Figures~\ref{fig:iin_C4_vs_c1_appendix} and~\ref{fig:gamma_C4_vs_c1_appendix}
depict the comparison of \ac{IIN} and $\Gamma$ measurements between sources in the
C4 catalogue and those in C1. For \ac{IIN}, the agreement is notably good, supported
by the mean and median values of the difference. Regarding $\Gamma$, despite a
larger observed scatter, the mean and median values of the difference suggest a
high level of agreement between measurements in the two catalogues.

Figure~\ref{fig:C4_vs_c2_pl_appendix} depicts a comparison of the $f_{\rm X}$,
\ac{IIN}, and $\Gamma$ measurements for sources included the C4 catalogue, where the
\zpltb{} model is utilised for spectral fitting (i.e. \acp{AGN}), with corresponding
measurements from C2, employing a power-law model for spectral fitting. The
$f_{\rm X}$ calculations exhibit a high level of agreement between the two
catalogues, evident from the mean and median differences and the scatter.
Similar consistency is observed in the comparison of photon index measurements.
Regarding \ac{IIN}, most sources show good agreement. A small fraction (${\sim}
2\%$) have \ac{IIN} values $<1$ in the C4 catalogue but ${\sim} 1$ in the C2
catalogue. These discrepancies in \ac{IIN} fits correspond to the minority of cases
where $f_x$ and/or $\Gamma$ differ.

Figure~\ref{fig:C4_vs_c2_bb_appendix} compares the $f_{\rm X}$ and \ac{IIN}
calculations for sources in the C4 catalogue, where their X-ray spectra are
fitted with a \bbpl{} model (i.e., \acp{XRB}) -- the absorbed sum of a blackbody and
a power-law component -- to those in the C2 catalogue, where spectra are
modeled using a single absorbed blackbody. The $f_{\rm X}$ measurements for
\acp{XRB} in the C4 catalogue appear to be higher by $0.2$~dex (on a logarithmic
scale) compared to those in the C2 catalogue. Conversely, the comparison of the
\ac{IIN} parameter reveals a good agreement between the measurements from the two
catalogues. We observe that the $p$-values are elevated in the C4 dataset
compared to the C2 catalogue, suggesting enhanced conformity in the former.
This outcome aligns with expectations, given the higher number of free
parameters in the model applied to the C4 dataset. We also find a quite strong
correlation between the difference in the $f_{\rm X}$ values and the difference
in the $p$-values between the two catalogues. Utilizing Spearman correlation
analysis, we calculated a correlation coefficient of 0.48, with a $p$-value of
$3.8 \times 10^{-12}$. This analysis indicates that the power-law component, in
conjunction with the blackbody, is a more appropriate model for fitting an
\ac{XRB} spectrum than a blackbody alone, as expected.

\FloatBarrier
\clearpage

\section{Catalogue column description}

The four catalogues described in this paper correspond to four distinct FITS
binary tables. Here we provide an overview of the information contained in the
catalogues. Each row in the catalogues corresponds to a unique identifier given
by either ${\tt DETID}$ (C1, C3), or ${\tt SRCID}$ (C2, C3), or a mixture of
both (C4). For each unique identifier, we report on the background-model
dependent camera net counts, camera usage, model parameter point estimate
values, and supplementary information (such as degrees of freedom, $p$-value,
and flag). Here we provide an overview of the naming convention used while the
catalogues themselves and additional documentation are available on-line:
\begin{itemize}
  \item C1: \url{https://zenodo.org/records/15193427},
  \item C2: \url{https://zenodo.org/records/15193545},
  \item C3: \url{https://zenodo.org/records/15267215},
  \item C4: \url{https://zenodo.org/records/15195730}.
\end{itemize}

The naming convention used for the camera counts and usage is as follows:
\begin{itemize}
  \item ${\tt pn\_cts}$      -- \pn{} total counts in the source extraction area,
  \item ${\tt pn\_bkgcts}$   -- \pn{} counts in the background extraction area,
  \item ${\tt pn\_netcts}$   -- \pn{} net counts in the source extraction area,
  \item ${\tt pn\_exp}$      -- \pn{} exposure time in seconds,
  \item ${\tt mos\_cts}$     -- \mos{} total counts in the source extraction area,
  \item ${\tt mos\_bkgcts}$  -- \mos{} counts in the background extraction area,
  \item ${\tt mos\_netcts}$  -- \mos{} net counts in the source extraction area,
  \item ${\tt mos\_exp}$     -- \mos{} exposure time in seconds,
  \item ${\tt det\_there}$   -- which cameras provide a spectrum 0: \pn{}; 1: \mos{}; 2: \pn{} and \mos{},
  \item ${\tt det\_use}$     -- which cameras are used for the fitting 0: \pn{}; 1: \mos{}; 2: \pn{} and \mos{}\@.
\end{itemize}

Each model is composed of one or more parameters. We use the following naming
convention for the parameters:
\begin{itemize}
  \item ${\tt lgflux}$    -- base-10 logarithm of the 0.2-12 keV flux in ${\rm erg/cm^2/s}$,
  \item ${\tt logNH}$     -- base-10 logarithm of the neutral hydrogen column density in ${\rm cm^{-2}}$,
  \item ${\tt PhoIndex}$  -- power-law photon index,
  \item ${\tt IIN}$       -- inter-instrument normalisation defined as \mos{}/\pn{},
  \item ${\tt kT}$        -- blackbody/APEC/bremsstrahlung plasma temperature in keV,
  \item ${\tt logNorm}$   -- base-10 logarithm of the power-law normalisation in ${\rm photons/keV/cm^2/s}$ (C3 only).
\end{itemize}

For each model parameter, we report the median and the mode from the posterior
distribution. For example, the median flux of the C1 power-law model is given by
${\tt lgflux\_med}$. We use the following naming convention for reporting the
median and the mode of each model parameter:
\begin{itemize}
  \item ${\tt med}$       -- median of the posterior distribution,
  \item ${\tt med\_min}$  --  5 per cent percentile of the posterior distribution,
  \item ${\tt med\_max}$  -- 95 per cent percentile of the posterior distribution,
  \item ${\tt mod}$       -- mode of the posterior distribution,
  \item ${\tt mod\_min}$  -- lower limit of the narrowest 90 per cent interval of the posterior distribution,
  \item ${\tt mod\_max}$  -- upper limit of the narrowest 90 per cent interval of the posterior distribution.
\end{itemize}

Finally, the following additional columns are provided in order to estimate the
quality of the model fit:
\begin{itemize}
  \item ${\tt dof}$           -- degrees of freedom of the model fit,
  \item ${\tt pvalue}$        -- KS $p$-value of the source+background fit,
  \item ${\tt pval\_bg\_pn}$  -- $\chi^2$ $p$-value of the background fit for \pn{},
  \item ${\tt pval\_bg\_mos}$ -- $\chi^2$ $p$-value of the background fit for \mos{},
  \item ${\tt flag}$          -- quality flag (described below).
\end{itemize}
The flag values 1 (no valid background counts), 2 (no valid net counts), 3
(unacceptable background fit) for C1, C2  and C3 are assigned to each spectrum
as they appear in its processing, and each of them exclude the spectrum from
further processing. The values assigned to each detection/source in the
catalogues correspond to the highest value reached (i.e.\ the spectrum that
reached the furthest processing stage) by the contributing \pn{} and \mos{}
spectra (or the value of the only spectrum that contributes).

For the catalogues C2 and C4, the model-dependent columns are suffixed by the
name of the model e.g. ${\tt lgflux\_med\_pl}$ refers to the median of the flux
of the power-law model, and ${\tt pvalue\_pl}$ refers to the corresponding
$p$-value. The model names used are:
\begin{itemize}
  \item ${\tt pl}$            -- power-law model (C2 only),
  \item ${\tt bb}$            -- blackbody model (C2 only),
  \item ${\tt zpltb}$         -- redshifted power-law model (C4 only),
  \item ${\tt apec\_single}$  -- APEC model (C4 only),
  \item ${\tt bbpl}$          -- blackbody + power-law model (C4 only),
  \item ${\tt bremss}$        -- bremsstrahlung model (C4 only).
\end{itemize}

For C4, we provide the following additional columns:
\begin{itemize}
  \item ${\tt D6}$             -- entry flag 1: detection is from C1; 2: source is from C2,
  \item ${\tt classification}$ -- source classification according to \citet{Tranin2022}. 0: \ac{AGN}; 1: star; 2: \ac{XRB}; 3: \ac{CV},
  \item ${\tt zbest}$          -- best available \ac{AGN} redshift, defined as the spectroscopic redshift if available, else photometric redshift.
\end{itemize}

\begin{acronym}
  \acro{AGN}{active galactic nucleus}
  \acro{BXA}{Bayesian X-ray Analysis}
  \acro{CSC}{Chandra Source Catalog}
  \acro{CV}{cataclysmic variable}
  \acro{GoF}{goodness of fit}
  \acro{IIN}{inter-instrument normalisation}
  \acro{SAS}{Science Analysis System}
  \acro{SOC}{Science Operations Centre}
  \acro{S/N}{signal-to-noise ratio}
  \acro{XRB}{X-ray binary}
  \acroplural{AGN}[AGNs]{active galactic nuclei}
  \acroplural{CV}[CVs]{cataclysmic variables}
  \acroplural{XRB}[XRBs]{X-ray binaries}
\end{acronym}

\end{document}